\documentclass[twocolumn,10pt]{IEEEtran}
\pdfoutput=1
\usepackage{comment}
\includecomment{toexclude}
\usepackage{amsmath,amsfonts,amssymb}
\usepackage{cite}
\usepackage{verbatim}
\usepackage{bm}
\usepackage{algorithm}
\usepackage{xcolor}
\usepackage[pdftex]{graphicx}
\usepackage{epstopdf}

\newcommand{\Yt}{\mathbf{Y}}

\newcommand{\st}{\mathbf{s}}
\newcommand{\sfd}{\tilde{\mathbf{s}}}
\newcommand{\sfi}{\tilde{s}}
\newcommand{\xt}{\mathbf{x}}

\newcommand{\Rt}{\mathbf{R}}

\newcommand{\hfi}{\tilde{h}}
\newcommand{\Ht}{\mathbf{H}}

\newcommand{\F}{\mathbf{F}}
\newcommand{\A}{\mathbf{A}}
\newcommand{\av}{\mathbf{a}}
\newcommand{\f}{\mathbf{f}}
\newcommand{\D}{\mathbf{D}}

\newcommand{\Pc}{\mathbf{P}}
\newcommand{\p}{\mathbf{p}}

\newcommand{\q}{\mathbf{q}}

\newcommand{\I}{\mathbf{I}}

\newcommand{\w}{\mathbf{w}}
\newcommand{\diag}{\text{diag}}

\newcommand{\C}{\mathbf{C}}
\newcommand{\Cq}{\widetilde{\mathbf{C}}}

\newcommand{\df}{\mathbf{d}}

\newcommand{\U}{\mathbf{U}}
\newcommand{\V}{\mathbf{V}}
\newcommand{\vt}{\mathbf{v}}

\newtheorem{theorem}{Theorem}

\title{Multibeam for Joint Communication and Sensing Using Steerable Analog Antenna Arrays}

\author{{J. Andrew Zhang,~\IEEEmembership{Senior~Member,~IEEE}, Xiaojing Huang,~\IEEEmembership{Senior~Member,~IEEE}, Y. Jay Guo,~\IEEEmembership{Fellow,~IEEE}, Jinhong Yuan,~\IEEEmembership{Fellow,~IEEE} and Robert W. Heath Jr,~\IEEEmembership{Fellow,~IEEE}} 
\thanks{J. Andrew Zhang, Xiaojing Huang and Y. Jay Guo are with University of Technology Sydney, Global Big Data Technologies Centre (GBDTC), Australia. Email:\{Andrew.Zhang; Xiaojing.Huang; Jay.Guo\}@uts.edu.au}
\thanks{J. Yuan is with the University of New South Wales, Australia. Email: j.yuan@unsw.edu.au.}
\thanks{Robert W. Heath Jr is with the University of Texas at Austin, Austin, TX 78712, USA. Email: rheath@utexas.edu.}
\thanks{The results in this paper was partially presented in our conference paper in IEEE Vehicular Technology Conference, 2017, Sydney, Australia.} 
}

\begin{document}

\maketitle

\begin{abstract}
Beamforming has great potential for joint communication and sensing (JCAS), which is becoming a demanding feature on many emerging platforms such as unmanned aerial vehicles and smart cars. Although beamforming has been extensively studied for communication and radar sensing respectively, its application in the joint system is not straightforward due to different beamforming requirements by communication and sensing. In this paper, we propose a novel multibeam framework using steerable analog antenna arrays, which allows seamless integration of communication and sensing. Different to conventional JCAS schemes that support JCAS using a single beam, our framework is based on the key innovation of multibeam technology: providing fixed subbeam for communication and packet-varying scanning subbeam for sensing, simultaneously from a single transmitting array. We provide a system architecture and protocols for the proposed framework, complying well with modern packet communication systems with multicarrier modulation. We also propose low-complexity and effective multibeam design and generation methods, which offer great flexibility in meeting different communication and sensing requirements. We further develop sensing parameter estimation algorithms using conventional digital Fourier transform and 1D compressive sensing techniques, matching well with the multibeam framework. Simulation results are provided and validate the effectiveness of our proposed framework, beamforming design methods and the sensing algorithms.
\end{abstract}

\begin{IEEEkeywords}
Joint Communication and Sensing, Beamforming, Multibeam, Compressive Sensing.
\end{IEEEkeywords}

\section{Introduction}\label{sec:intro}

There have been increasing demands for systems with both communication and (radar) sensing capabilities, on emerging platforms such as unmanned aerial vehicles and smart cars \cite{Han13, KumariCPH17}. Radio sensing here is referred to as information retrieval and harvesting, based on estimating the position, speed and feature signal of objects in the environment that radio signal covers. Rooting from traditional radar technology, radio sensing is evolving with significantly expanded scope and applications \cite{Yousefi17}. Instead of having two separate systems, it is possible to develop joint communication and sensing (JCAS) techniques to integrate the two functions into one by sharing hardware and signal processing modules, and achieve immediate benefits of reduced cost, size, weight, and better spectrum efficiency. An integrated system will also benefit from mutual sharing of information for improved performance, e.g., using sensed environment knowledge to assist beamforming (BF) design \cite{Gonza16}.

Millimetre wave (mmWave) provides a great platform for integrating these two functions into one module thanks to its large bandwidth and small antennas \cite{Choi16}. With large bandwidth, it provides potentially a high data rate for communication, high resolution for radar operation, and low latency for information exchange and sharing between the two modules. For mmWave, antenna array and steerable beam need to be used to overcome large propagation attenuation and to estimate signal direction. Beam-steering can be realized either via a full digital or analog array. A full digital array for mmWave systems is very expensive, while analog array is a practical and cost-effective solution. Commercially available mmWave phased array, e.g., $16\times 16$ $60$ GHz millimetre wave arrays of size $22\times 22$ cm$^2$ \cite{Zihir15}, provides great potential for JCAS, particularly for small platform.

Although BF for communication and radar sensing respectively has been studied extensively \cite{vanveen88, Sturm11}, its applications in JCAS is not straightforward. The main challenge is that communication and sensing have different requirements for BF. Radio systems operating at high frequency bands confront large propagation loss. In this case, sensing requires time-varying directional scanning beams, while communication requires stable and accurately-pointed beams to achieve large BF gain. Existing research mostly considers the scheme of using a single beam for communication and sensing \cite{Kumari15, Yan16, Va16, KumariCPH17}. The direction of sensing in this case is limited to the direction of the communication node.

In this paper, exploring novel \emph{multibeam} technology we present a comprehensive framework that enables low-cost and efficient implementation of JCAS on even small and portable platforms, with the use of two analog antenna arrays. Our proposed approach, as a first initiative of using multibeam technology in JCAS, is an innovation significantly different to existing ones \cite{Kumari15, Yan16, Va16, KumariCPH17}. A multibeam is defined as a BF waveform with two or more mainlobes (we call them \textit{subbeams}) generated by a single analog antenna array at a time. Some subbeams, fixed in direction, support communication as well as sensing in the communication direction(s), and other subbeams support direction-varying scan over different communication packet durations. The proposed framework enables seamless integration of sensing into a (point-to-point) communication system with multicarrier modulation and packet transmission. This paper provides a system architecture, operation protocols, a BF design methodology, multibeam generation and updating algorithms, and sensing algorithms, to achieve large field-of-view (range in directions), flexible and accurate sensing with controllable and insignificant compromise on communication performance.

Our main contributions in this paper are as follows, with a focus on multibeam with two subbeams: 
\begin{itemize}
\item We propose a system architecture and protocols for implementing multibeam JCAS technology, with the use of two analog arrays. The two arrays, spatially widely separated, are introduced  mainly to suppress leakage signal from the transmitter to receiver, so that the receiver can work all the time, transiting between communication and sensing modes. The proposed framework fully complies with a conventional time division duplex (TDD) communication system and reuses the TDD timeslot. This is detailed in Section \ref{sec-problem};
\item We provide a BF design methodology to enable the integration of beam-scanning based sensing with fixed-beam based communication functions. We also present a generalized Least Squares (LS) BF solution for generating beams with desired shape. This is discussed in Section \ref{sec-bfdesigns}; 
\item Detailed methods for generating and updating multibeam BF vectors are developed. The methods offer flexibility in meeting different and time-varying requirements for communication and sensing, capability in constructively combining communication and sensing subbeams for communication purpose, and simplicity in updating BF vectors. The details are presented in Section \ref{sec-multibeam};
\item We formulated the signal processing and sensing parameter estimation problems, and investigated sensing algorithms for estimating these parameters. In particular, we developed novel 1D Compressive Sensing (CS)-based solutions, which well-match the multibeam framework and can work efficiently in, e.g., smart cars and commercial UAV networks. The work is presented in Section \ref{sec-sensing}.
\end{itemize}

Notations: $(\cdot)^H$, $(\cdot)^T$ and $(\cdot)^c$ denote the Hermitian transpose, transpose and conjugate of a matrix/vecor, respectively. $|\cdots|$ denotes the element-wise absolute value, $(\A)_{n,m}$ denotes the $(n,m)$-th element of the matrix $\A$, $(\A)_{\cdot,m}$ denotes the m-th column of $\A$, $\{a_n\}$ denotes a vector with elements $a_n$, $\diag\{a_n\}$ denotes a diagonal matrix with diagonal elements $a_n$.
 
\section{Proposed System Architecture, Protocol and Signal Model}\label{sec-problem}

We consider a system where two nodes perform two-way point-to-point communication in time division duplex (TDD) mode, and simultaneously sensing the environment to determine locations and speed of nearby objects. Using TDD allows better hardware sharing and avoids complex synchronization between two-way transmissions, compared to frequency division duplex. Each node uses two spatially (widely) separated steerable antenna arrays. The primary goal for using two arrays is to suppress the leakage from transmitter to receiver, as the receiver always needs to be in operation, time-switched between sensing and communication. One array is dedicated to the receiver, and the other can be shared by transmitter and receiver through time division. We consider orthogonal frequency division multiplexing (OFDM) here for its popularity in modern communication systems, and its strong potential for sensing \cite{Sturm11}. The proposed framework can be extended to other packet-based communication systems.

\subsection{System Architecture and Protocol}
\begin{figure}[tb]
\centering
\includegraphics[width=\columnwidth]{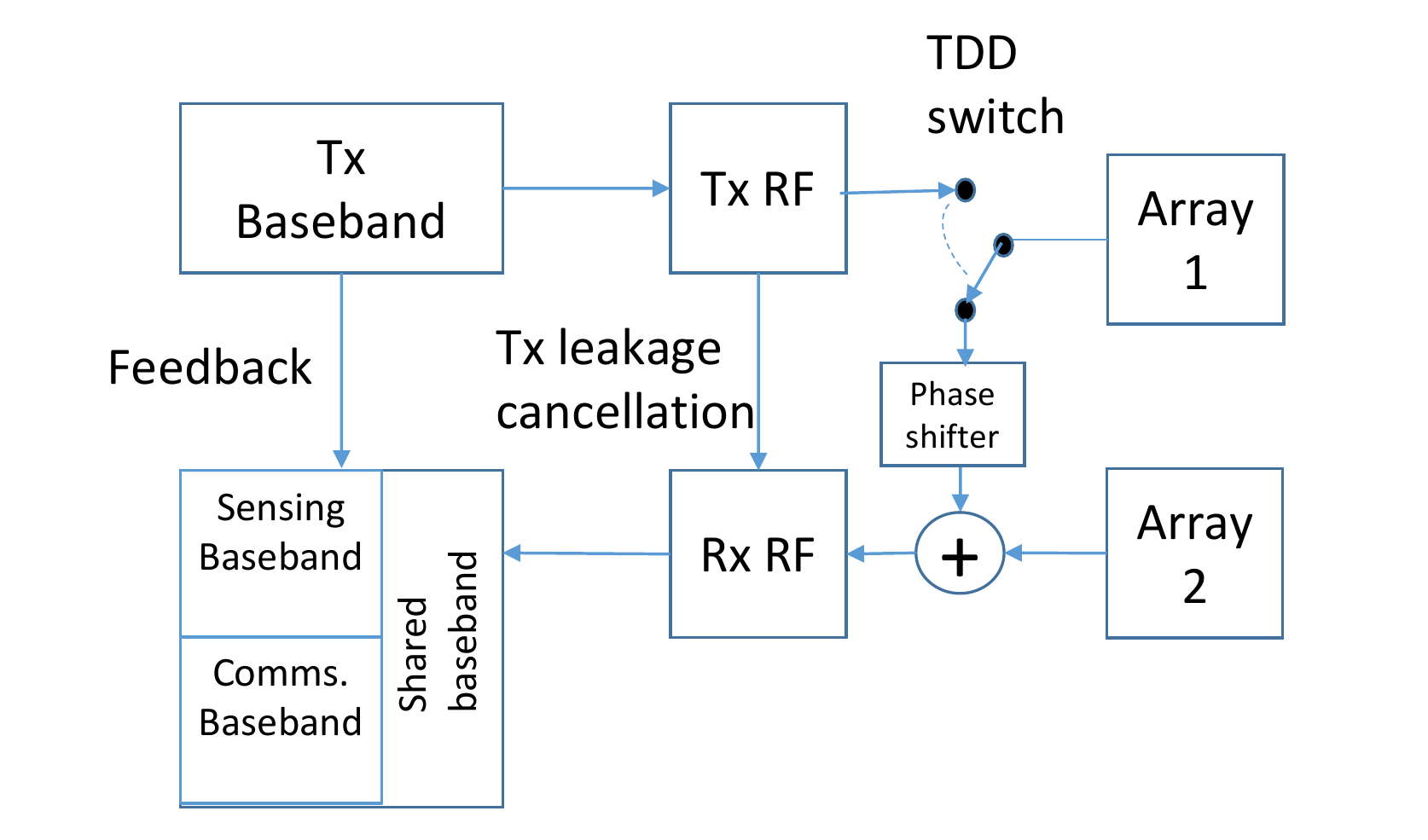}
\caption{Block diagram of the basic transceiver that uses two analog arrays. The two arrays are mainly used for suppressing leakage signal from the transmitter to the receiver so that the receiver can operate all the time.}
\label{fig-trx}
\end{figure}

Fig. \ref{fig-trx} shows the diagram of the proposed transceiver. The transmitter baseband module is common to communication and sensing. The baseband signal is sent to the transmitter radio frequency  (RF) frontend, and radiated through Array 1. Array 1 is primarily used for the transmitter and can be optionally connected to the receiver through an electronic switch and a digitally controlled phase shifter. The transmitter RF signal after power amplifier can also be optionally fed to the receiver RF for cancelling leakage signal from the transmitter.

At the receiver, Array 2 is always connected to the RF frontend, and Array 1 is only connected when  the node itself is not transmitting and only receiving signal from the other node. Adding Array 1 can ideally double the BF gain in this case, where signals  from the two arrays are constructively added together by a combiner before the RF module. Communication and sensing at the receiver baseband share some signal processing modules such as discrete Fourier transform (DFT), channel estimation and equalization, but are largely different in subsequent processing. The receiver baseband also accepts feedback from the transmitter baseband, mainly for getting a clear sensing signal by removing the data symbols from the received signal.

\begin{figure}[t]
\centering
\includegraphics[width=\columnwidth]{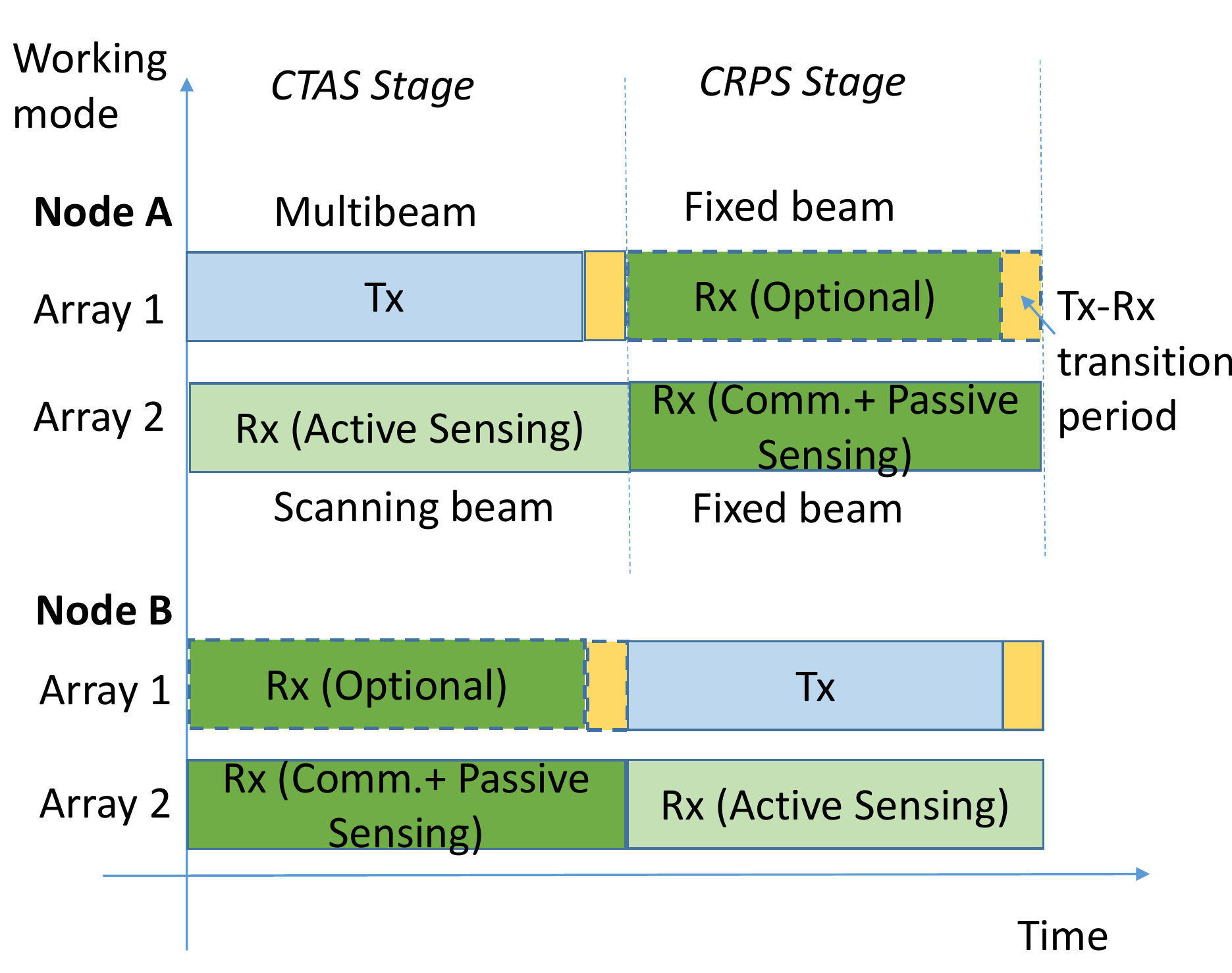}
\caption{Procedure and protocol of communication and sensing in a point-to-point connection scenario. Communication is in the TDD mode.}
\label{fig-cycle}
\end{figure}

Fig. \ref{fig-cycle} illustrates the proposed procedure and protocol for JCAS between two nodes A and B. The two nodes communicate in the TDD mode, and the transmitted signal from each node is used for both communication and sensing. For each node one complete cycle includes two stages: \emph{Communication Transmission and Active Sensing (CTAS)}, and \emph{Communication Reception and Passive Sensing (CRPS)}. We refer active and passive sensing to the cases where sensing signal is transmitted by the node itself and by other nodes, respectively. There are two major differences between them: 1) In the former the transmitter and receiver can be synchronized in clock and hence the measured time delay is absolute; while in the latter, the measurement is typically relative due to the lack of synchronization; 2) The transmitted signal is known to the receiver in the former, while it is typically unknown, but may be decoded and reconstructed, in the latter. In addition, the sensed environment can be different due to different propagations: in active sensing, most received signals are reflected ones and the sensing results are more of a radio imaging of the environment that the node confronts; while in passive sensing, most received signals are refracted ones and they also contain the transmitter's information.
   
From Node A's viewpoint, we now describe the detailed implementation in the two stages. In the CTAS stage of Node A, when Node B is in the CRPS stage, Node A's transmitter uses Array 1 to generate a multibeam, with one subbeam pointing to Node B and the other subbeam adapting to the sensing requirement. During this stage Array 2 of Node A is used for sensing only. It typically forms one narrow single-beam and scans the direction corresponding to the transmitter scanning beam. At the end of the CTAS stage, there is a short transition period between transmission and reception, as usually exists in a TDD transceiver. This period also serves as a guard interval for Array 2, such that the reflected signals from its own transmitter will be separated from the received signals from Node B's transmitter in the following CRPS stage.

In Node A's CRPS stage when Node B is in its CTAS stage, Array 2, as well as Array 1 optionally through a switch, work in the receiving mode, and their signals are combined and processed, primarily for communication, and optionally for passive sensing. Sensing in this case uses transmitted signal from Node B. The two arrays in this stage can be treated equally, and optimized jointly to achieve best results for communication, as well as passive sensing. 

This protocol reuses the TDD frame structure for communication and sensing, i.e., downlink and uplink sensing uses downlink and uplink slots, respectively. The TDD frame structure impacts the continuity of sensing, and if possible, it can be optimized by jointly considering communication and sensing needs.

To make the system work, BF design, generation and updating of the multibeam, and the corresponding sensing and communication algorithms are critical problems to be solved. These problems will be addressed in the subsequent sections.

\subsection{Formulation of Signal Model}

We consider planar wavefront and uniform linear antenna arrays (ULA) with antenna elements equally spaced at half wavelength in this paper. For an array with $M$ antennas, the array response vector for an angle-of-departing (AoD) $\theta_t$ or angle-of-arrival (AoA) $\theta_r$ is given by
\begin{align}
\av(\theta)=[1,e^{j\pi \sin(\theta)},\cdots, e^{j\pi (M-1)\sin(\theta)}]^T 
\label{eq-avm}
\end{align}
where $\theta$ is for either $\theta_t$ or $\theta_r$.

In the OFDM system, let $N$ denote the number of subcarriers and $B$ the total bandwidth. We get the subcarrier interval $f_0=B/N$ and OFDM symbol period $T_s=N/B+T_p$ where $T_p$ is the period of cyclic prefix. The baseband signal in the transmitter can then be represented as
\begin{align}
\st=\F^H\sfd,
\end{align}
where $\sfd$ is the $N\times 1$ data symbol vector, $\F$ is the DFT matrix, and $\st$ is the time domain signal. Let $s(t)$ denote the time-domain baseband signal with cyclic prefix appended. The signal transmitted from the antenna array is
\begin{align}
\xt(t)=s(t)\w_t,
\end{align}
where $\w_t$ is the transmitter BF vector. 

We can use a common channel model for communication and sensing, although the actual propagation mechanism could be very different. Consider $L$ multipath signals with AoDs  $\theta_{t,\ell}$ and AoAs $\theta_{r,\ell}$. For the simplicity of notation, we assume that transmitter and receiver arrays have the same number of antennas $M$. The results in this paper can be straightforwardly extended to arrays with different number of antennas and antenna intervals. The time-varying physical channels between the transmitting and receiving antennas can then be represented as
\begin{align}
\Ht=\sum_{\ell=1}^L b_\ell \delta(t-\tau_{\ell})e^{j2\pi f_{D,\ell} t} \av(\theta_{r,\ell})\av^T(\theta_{t,\ell}),
\label{eq-Ht}
\end{align}
where for the $\ell$-th multipath, {$b_{\ell}$ is its amplitude of complex value accounting for both signal attenuation and initial phase difference},  $\tau_{\ell}$ is the propagation delay, and $f_{D,\ell}=10^{-8}v_sf_c/3$ is the associated Doppler frequency that causes time varying of $\Ht$. Here $v_s$ is the relative moving speed, $f_c$ is the carrier frequency. {Note that for communications, we generally only need to know the elements in the matrix $\Ht$, and do not have to resolve the detailed channel structure and estimate the detailed parameters as shown in (\ref{eq-Ht}). When the multipath is sparse,  estimating these underlying parameters directly  can be simpler than estimating the matrix elements \cite{Ayach14}. For active sensing, of course the channels could be different for communication and sensing.} 

Let the receiver BF vector be $\w_r$. The signal arriving at a receiver array for either sensing or communication can be represented as
\begin{align*}
&y(t)\notag\\
&= \sum_{\ell=1}^L b_\ell e^{j2\pi f_{D,\ell} t} (\w_r^T\av(\theta_{r,\ell}))\cdot(\av^T(\theta_{t,\ell})\w_t)s(t-\tau_{\ell}) + z(t)
\end{align*}
where $z(t)$ is the additive white Gaussion noise (AWGN).

\section{Beamforming Design}\label{sec-bfdesigns}
We assume a packet-based communication system where each packet contains many OFDM symbols, and each node sends multiple packets to the other node in one cycle. In such a system, there are two essential requirements for JCAS: (1) having stable and high-gain beams for communication to avoid complex channel tracking within a packet; and (2) having direction-varying scanning beams to enable sensing over a large area of interest. Next we propose beamforming design approach that meets both requirements.

\subsection{Basic Constraints for BF Vectors}\label{sec-principle}
These two requirements impose two basic constraints for BF vectors $\w_t$ and $\w_r$:
\begin{itemize}
\item $\w_t$ needs to be fixed during at least one packet to make the channel stable for communication (This can be relaxed if powerful channel tracking is available, which is not easy to implement);
\item For sensing $\w_r$ needs to be fixed during at least one OFDM symbol so that information symbols can be removed from the received signal; for communication, it needs to be fixed over the whole packet.
\end{itemize}
These two constraints can generally be satisfied in a typical mmWave system. Consider an OFDM system with carrier frequency $f_c=24$GHz, bandwidth $B=100$MHz and $N=128$ subcarriers. The OFDM symbol period is about $T_s=1.6u$s (assuming $T_p=N/(4B)$). Within 1ms which approximately equals to the period of $625$ OFDM symbols, the moving distance is less than $5.6$cm for a relative speed smaller than $200$ km/h. Hence during this 1ms period, the channel parameters as shown in (\ref{eq-Ht}) can be assumed to be pseudo-static and unchanged. In this case, $f_{D,\ell}\leq 4.5$kHz and $f_{D,\ell}T_s<0.0072$, and the Doppler phase shift can be regarded as unchanged over multiple $T_s$s.  For more details on the impact of OFDM parameters on sensing, the readers are referred to \cite{Sturm11}.

In addition, to ensure sufficient power for communication, at least one directional beam should point to the target node, similarly for the receiver in the CRPS stage. We also want the transmitter scanning beam to be directional, so that sufficient energy could be reflected for sensing. Their directivity can assist the coarse estimation of AoDs, too. 

\subsection{Proposed BF Design}\label{sec-bfdesign}
Considering the constraints above, our proposed BF design based on multibeam is as follows.

(1) \textit{Fix $\w_t$ during one packet}. Design $\w_t$ to generate multibeam consisting of one directional communication subbeam and another directional subbeam for scanning and sensing. The communication subbeam always points to the target node for communication, and the sensing subbeam points to one or more directions at a time for sensing. The sensing subbeam scans the whole directions of interest over $N_t$ packet periods. The width and gain of the scanning subbeam can be different from the communication subbeam which also senses the fixed direction pointing at the communication node;

(2) \textit{For sensing, design $N_r$ different $\w_r$, and apply the $k$-th, $k\in[1, N_r] $, $\w_r$ to $N_d$ OFDM symbols with indexes $k, k+N_r,\cdots,k+(N_d-1)N_r$}. This interleaved variation of $\w_r$ is mainly to enable better estimation of Doppler frequency, which is very small and needs to be measured over a long period. The difference between $\w_r$ allows wider total scanning directions and can also potentially allow accurate estimation of the AoAs. Each of the $N_r$ $\w_r$s can be either random vectors, e.g., fulfilling compressive sensing requirement, or vectors generating directional beams pointing to one or more desired scanning directions. Hence the receiver BF can be either single beam or multibeam. In Section \ref{sec-rxbf}, we will provide an exemplified receiver BF design, adapting to a two-beam transmitter beamforming;

(3) \textit{For communication, $\w_r$ for the two arrays shall be ideally jointly designed to achieve the optimal results}. Due to their difference in orientation and location, this could be very challenging. A suboptimal approach is to design the beamforming for each array independently and then apply a single phase shift to the signal from Array 1 to ensure a constructive combining of signals from the two arrays. If no passive sensing is required, each array can simply generate a single beam pointing to the transmitting node; otherwise, $\w_r$ needs to be fixed during one packet to keep channel fixed, and hence no scanning is implemented (Detailed design and implementation will not be discussed in this paper).

Therefore we have assumed that each packet has $N_rN_d$ consecutive OFDM symbols, indexed from $1, \cdots, N_rN_d$, and one complete scanning requires $N_t$ packets and $N_tN_rN_d$ OFDM symbols. The width and gain for sensing and communication subbeams can be designed to adapt to these parameters, and vice verse. The values of $N_t$, $N_r$, and $N_d$ could have notable impact on the sensing performance, and hence they can be optimized in favour of sensing, while meeting the communication requirements. In general, $N_t$ depends on the ratio between channel correlation time and the packet period, and affects the sensing ranges in distance and angle. The product of $N_rN_d$ depends on the number of available OFDM blocks in one packet, and their respective values affect sensing ranges too. We will discuss the design of BF waveforms with given values of these parameters in Section \ref{sec-multibeam}. Detailed parameter optimization is beyond the scope of this paper.

\subsection{Least Squares BF Solution with Desired Array Response}\label{sec-gls}
The preceding proposed BF design demands algorithms that can flexibly generate BF vectors achieving desired array response. Assume that the desired array response is known, we now provide a generalized least squares (LS) beamforming design method, show its optimality, and extend it to the case when only the magnitude of the desired response is known. This method forms the basis for the multibeam generation and updating algorithms as will be detailed in Section \ref{sec-multibeam}.

\subsubsection{Generalized LS Solution and its Optimality}
It is well known that the LS solution for a conventional BF design problem 
\begin{align}
\A \w=\vt
\label{eq-vtls}
\end{align}
is given by $\w_{\text{LS}}=\A^\dagger\vt$, where $\w$ is the BF vector, $\A=[\av_1,\cdots,\av_K]^T$ is the array response matrix at $K$ specified directions, $\vt=[v_1,\cdots,v_K]^T$ is the desired array response at these directions , and $\A^\dagger=(\A^*\A)^{-1}\A^*$ denote the pseudo-inverse of $\A$. {Here, the parameter $K$ does not directly depends on the number of multipath $L$.  Since the number of antennas in an array, $M$, determines the beamwidth and then its spatial resolution, the value of $K$ shall be selected to be proportional to $M$, for example, $K$ equals to 4 to 6 times of $M$. A larger $K$ could lead to better granularity in specifying the array response, but will also lead to higher computational complexity. Since the BF weight vector for generating the basic BF waveform can typically be pre-computed and stored in the system, the computational complexity is not a real problem. } 

The solution $\w_{\text{LS}}=\A^\dagger\vt$ to (\ref{eq-vtls}), however, does not apply a power constraint to $\w$, which cannot validate the optimality when power constraint is actually necessary for both transmitter and receiver beamforming.

Here we formulate a generalized LS problem for beamforming design, derive its solution and show its optimality. We consider a constrained and weighted LS problem
\begin{align}
&\min_{\w,c_s} \parallel \D(c_s\A \w-\vt) \parallel_2^2,\notag\\
&\text{s.t.}\ \w^H\w=1,
\label{eq-ls}
\end{align}
where $\D$ is a real diagonal weighting matrix with diagonal elements' mean power normalized to 1, and $c_s$ is a complex scalar to be determined. The pre-chosen matrix $\D$ can be used to impose different accuracy requirements on different segments of the generated BF waveform. Assume that $\A^\dagger$ exists, which is easily achievable by spatial sampling design.

The optimal $c_s$ for (\ref{eq-ls}) can be obtained via computing the Wirtinger derivative of $\parallel \D(c_s\A \w-\vt) \parallel_2^2$ with respect to $c_s$ and letting the derivative be zero. This leads to
\begin{align}
c_s=\frac{(\D\A\w)^H\D\vt}{\parallel\D\A\w\parallel_2^2}.
\label{eq-a}
\end{align}

Substituting (\ref{eq-a}) to (\ref{eq-ls}), the original optimization problem becomes
\begin{align}
&\max_{\w} \frac{\w^H\Rt_1\w}{\w^H\Rt_2\w},\notag\\
&\text{s.t.}\ \w^H\w=1,
\label{eq-ls3}
\end{align}
where $\Rt_1=((\D\A)^H\D\vt) ((\D\A)^H\D\vt)^H$, and $\Rt_2=(\D\A)^H(\D\A)$.

The objective function in (\ref{eq-ls3}) is known as \textit{Rayleigh quotient}. 
It is known to be maximized when $\w$ equals to the eigenvector $\w_{\text{opt}}$ corresponding to the maximum eigenvalue in the following generalized eigenvalue problem
\begin{align}
\Rt_1\w=v\Rt_2\w.
\label{eq-gv}
\end{align}
The eigenvector has a norm of $1$, and hence the constraint in (\ref{eq-ls3}) is automatically satisfied. The maximal generalized eigenvalue $v_{\text{max}}$ is given by
\begin{align}
v_{\text{max}}=\frac{\w_{\text{opt}}^H\Rt_1\w_{\text{opt}}}{\w^H_{\text{opt}}\Rt_2\w_{\text{opt}}}.
\end{align}

Assume $\Rt_2$ is invertible. Noticing that $\Rt_1$ is the outer product of a vector $((\D\A)^H\D\vt)$ and its conjugate, via substituting $v_{\text{max}}$ to (\ref{eq-gv}) we can remove common scalars from the two sides of the resulted equation and obtain
\begin{align}
\w_{\text{opt}}=\alpha\,\Rt_2^{-1}(\D\A)^H\D\vt,
\end{align}
where
\begin{align}
\alpha\triangleq\frac{\w_{\text{opt}}^H\Rt_2\w_{\text{opt}}}{\w_{\text{opt}}^H(\D\A)^H\D\vt}
\end{align}
is a scalar that normalizes the power of $\w_{\text{opt}}$ to 1.

When $\D=\I$ we have
\begin{align}
\w_{\text{opt}}=\alpha\,\A^\dag\vt.
\label{eq-lsn}
\end{align}
Interestingly, the optimal solution is now a normalized  LS solution. Hence we can get the following theorem:
\begin{theorem}
The optimal solution to the constrainted problem: $\min_{\w,c_s}\parallel (c_s\A \w-\vt) \parallel_2^2$, \text{s.t.} $\w^H\w=1$, is equal to the normalized least squares solution to $\min_\w\parallel \A \w-\vt \parallel_2^2$.
\end{theorem}

Theorem 1 indicates that the conventional normalized LS solution $\A^\dagger\vt$ is actually also the optimal LS solution when the power constraint $\w^*\w=1$ is required.

\subsubsection{Known Magnitude Only and Iterative LS Algorithm}
In general, we can only specify the desired magnitude of the elements in $\vt$, but not their phases. Let
\begin{align}
\vt=\D_v\p_v,
\end{align}
where $\D_v$ is a diagonal matrix with diagonal elements being the magnitude of the elements in $\vt$, and $\p_v$ is the phase vector for $\vt$. Actually, the phases provide $K$ degrees-of-freedom for minimizing the least square error $\parallel \A\A^\dagger\vt-\vt\parallel_2^2$. This can be further formulated as:
\begin{align}
\p_{v,\text{opt}}=\arg\min_{\p_v}\parallel (\A\A^\dagger-\I)\D_v\p_v\parallel_2^2
\end{align}
This optimization problem is not easy to solve since each element of $\p_v$ needs to be on a unit circle. 

The two-step iterative least squares (ILS) method in \cite{Shi05} provides a sub-optimal solution for $\w_{\text{opt}}$ by exploiting the freedom of choosing $\p_{v,\text{opt}}$. Although several further improvements to the algorithm were made, its convergence is not proven yet. Actually, we have found that the algorithm quite often converges to the same BF vector even when the given $\D_v$ is different, e.g., when taken from different length of a given curve. We will discuss detailed observations for the behaviour of this algorithm and propose how to choose appropriate initial $\D_v$ in Section \ref{sec-simu} for our problem.
 
\section{Generation and Updating of Multibeam BF Vectors}\label{sec-multibeam}

Based on the BF design principles and ILS BF generation algorithm in Section \ref{sec-bfdesigns}, we now provide detailed methods for generating and updating multibeam BF vectors. We consider a narrowband BF model with continuous, instead of quantized BF vector values. We will also assume that there is a dominating multipath for the communication link. 

We want to achieve the following goals for the multibeam design: 
\begin{enumerate}
\item[G1:] The multibeam can simultaneously meet communication and sensing requirements for various beamwidth and power levels, and suit signals with different packet parameters;
\item[G2:] Communication and sensing subbeams can be added up constructively to obtain an improved SNR for at least the communication link;
\item[G3:] BF vectors can be updated simply and rapidly to adapt to changed communication and sensing requirements in real time.
\end{enumerate}

The complete process includes the following steps, as will be detailed in each of the following subsections:
\begin{enumerate}
\item[S1.] Generate the \textit{basic reference BF waveform}s for communication and sensing (Pointing at zero degree) as will be discussed in Section \ref{sec-basicbf} ;
\item[S2.] Shift the reference waveforms to the desired directions using the \textit{displaced BF waveform} algorithm to be proposed in Section \ref{sec-displace};
\item[S3.] Combine the communication and scanning subbeams and obtain the normalized BF vector $\w_{t}$, using one of the methods to be proposed in Section \ref{sec-methods}. 
\end{enumerate}

In Section \ref{sec-rxbf}, we will also present an exemplified receiver BF design that corresponds to the transmitter multibeam. 

\subsection{Generating Basic Reference BF Waveform} \label{sec-basicbf}

The basic BF waveform for communication and sensing will be separately generated using the ILS algorithm, with possibly different beamwidth and shape.

This involves the determination of beamwidth, number of scanning directions and the shape of the BF waveform. The reference waveforms generally remain unchanged for a long period, and do not need to be regenerated if only scanning and communication directions change. They will be used as subbeam inputs to the subsequent steps and generate the final multibeam waveform. Note that the subbeams in the finally resulted multibeam may be slightly different from the desired shape.
 
We can determine the pointing directions and required beamwidth of the scanning subbeam, based on the desired scanning range and the values of $N_t$, $N_r$, and $N_d$. Wider beams can be used to cover a larger scanning range and/or complete scanning within a less number of packets. However, they generally lead to reduced sensing range due to reduced BF gain, as well as reduced angle resolution. A general practice is to let scanning directions to be spaced at 3dB beamwidth, which can be approximately calculated as $2\text{arcsin}(1.2/M)$ in radius for a ULA with $M$ omnidirectional antenna elements spaced at half wavelength.  

The shape of the BF waveform can be regulated by the input to the ILS algorithm via specifying the desired BF magnitude $\D_v$. To generate a beam with small sidelobes using the ILS algorithm, we found it effective to let $\D_v$ be the mainlobe of a BF radiation pattern, and setting the rest samples to be zeros, as will be shown in Fig. \ref{fig-refbeam}.

The power allocation between the communication and scanning subbeams depends on the required communication distances and the sensing ranges. The power allocation can be realized via either adjusting the beamwidth or distributing energy proportions between the two subbeams. The latter can be conveniently achieved as will be shown in Section \ref{sec-methods}.  There is always a compromise between the communication and sensing range requirements as the increase in one will generally cause reduction to the other.  

\subsection{BF Vectors for Displaced BF Waveform}\label{sec-displace}
In this subsection we show that a BF waveform displaced by an ``equivalent'' scanning direction can be obtained through element-wise multiplying the original BF vector by a phase shifting vector.

We first show that BF waveform displaced by an actual scanning direction cannot be obtained in a simple way. Consider an $M$-element ULA whose array response vector at the direction $\theta$ is given by $\av(\theta)$ in (\ref{eq-avm}). Let $\w_1$ be the current BF vector. The radiation pattern (BF value) at the direction $\theta$ is given by 
\begin{align}
r(\theta)=\av^T(\theta)\w_1.
\end{align}  
Now we want to use $\w_2$ to generate the same BF value at the direction $\theta+\delta$. This can be represented as
\begin{align}
r(\theta+\delta)=&\av^T(\theta+\delta)\w_2\notag\\
=&\av^T(\theta)(\D_\delta(\theta)\w_2)
\label{eq-w2}
\end{align}   
where $\D_\delta(\theta)$ is a diagonal matrix with the $m$-th diagonal element  equal to $\exp(j\pi m(\sin(\theta+\delta)-\sin(\theta)))$. It is clear that $\D_\delta(\theta)$ depends on $\theta$, and hence such BF waveform displacement cannot be achieved via multiplying a single phase shifting vector. 

Now let us consider an $M\times Q$ array response matrix $\A$ with quantized and equally spaced element values. For its $q$-th column,
\begin{align}
\av_q=[1,  \cdots, e^{j\pi mq\phi}, \cdots,e^{j\pi (M-1)q\phi}]^T,
\end{align}
where $\phi$ is the quantization step, $q\phi$ can be treated as an \textit{equivalent direction}, corresponding to $\sin(\theta)$ with $\theta$ being the actual direction. We can make $\phi$ sufficiently small, meeting desired pointing direction resolution. However, $\av_q$ is not equally spaced with respect to the actual directions. 

Now we want to generate displaced BF waveform with equivalent directions changing from $q\phi$ to $(q+\delta)\phi$. We can obtain the relationship between $\w_1$ and $\w_2$ as
\begin{align}
\w_2=(\D_\delta)^c\w_1,
\end{align} 
where $(\D_\delta)$ is an $M\times M$ diagonal matrix with $m$-th diagonal elment being $\exp(j\pi m\delta\phi))$. Since $\D_\delta$ is independent of $q$, the whole BF waveform can now be displaced by multiplying $(\D_\delta)^c$ to $\w_1$.

This suggests that we only need to pre-generate one BF vector $\w_1$ from a reference BF waveform with desired shape, then obtain any displaced BF waveform by multiplying a phase shifting vector to it. Note that the shape of the waveform is retained only in the equivalent directions, but not the actual directions (See the examples in Figs. \ref{fig-indi} and \ref{fig-multibeam}). However, for most applications, such distortions in BF waveform is acceptable, particularly given the benefits from the simplicity of this method.  

\subsection{Generating and Combining Communication and Sensing Subbeams}\label{sec-methods}

In this subsection, we present two methods for generating a multibeam that can achieve a good balance between preserving the desired shape of the BF waveform and the combined BF gain for communication. Method 1 first generates the communication and sensing subbeams separately, and combine them by introducing a vector-level phase shifting term. Method 2 generates the two subbeams jointly.

\subsubsection{Method 1: Separated Design + Vector-level Phase Alignment}

In this separated design, we first generate the communication and sensing beams separately and then combine them coherently. We split the transmitter BF vector as
\begin{align}
\w_t=\sqrt{\rho}\w_{t,c}+\sqrt{(1-\rho)}\w_{t,s},
\end{align} 
where $\w_{t,c}, \parallel\w_{t,c}\parallel_2^2=1$, and $\w_{t,s}, \parallel\w_{r,c}\parallel_2^2=1$ are primarily for generating beams for communication and sensing, respectively, the parameter $\rho$, $0\leq\rho\leq 1$ controls the energy distribution between the two BF vectors. The power of $\w_t$ is further normalized to 1. 

{The value of $\rho$ can be determined by jointly considering the communication and sensing requirement. Note that for active sensing, the path loss factor is about 4, while for communication and passive sensing, it is about 2. Hence the range of active sensing is approximately the square root of the communication distance, given that the received SNRs are the same. Noting this difference, we can then decide a proper value for $\rho$ to balance the communication and sensing requirements.}
	
For $\w_{t,c}$, we want to determine its value to achieve the transmitter BF gain as high as possible, depending on available channel knowledge. We consider two cases in this paper: (1) when the full channel matrix $\Ht$ is known, and (2) when the (estimated) dominating AoD is known. 

In the first case, let $\p_1$ and $\q_1$ be the left and right singular vectors of $\Ht$, corresponding to the maximum singular value. In order to maximize the received communication power at Node B, ideally we would like to let $\q_1^c$ and $\p_1^c$ be the transmitting and receiving BF vectors, respectively. However, letting $\w_t=\q_1^c$ is not always possible since $\w_{t,s}$ needs to vary over packets. Hence we let $\q_1^c$ to be the initial value of $\w_{t,c}$ in this case.

In the second case, we consider the approach of setting the initial value of $\w_{t,c}$ to be $\av^c(\theta_{t,c})$ to generate a BF pointing to the dominating AoD $\theta_{t,c}$. 

In either case, $\w_{t,c}$ can be further optimized from these initial values in order to, e.g., reduce the sidelobes, using the ILS method. Hence the final $\w_{t,c}$ can become slightly different.

Using case 2 $\w_{t,c}=\av^c(\theta_{t,c})$ as an example, we now propose a method for designing $\w_{t,s}$ to achieve coherent signal combination for the two subbeams for the communication link. This process can be directly applied to case 1 by substituting $\av^c(\theta_{t,c})$ with $\q_1^c$. 

According to the ILS solution in Section \ref{sec-gls}, we let $\p_{v,0}$ and $\bm{\beta}\triangleq\alpha\A^\dag \D_v\p_{v,0}$ be the output phase vector and the BF vector in the ILS algorithm, respectively. Let $g_c=\av^T_{\theta_{t,c}}\w_{t,c}$ and ${g_s}=\av^T_{\theta_{t,c}}\bm{\beta}$. Applying a vector-level adjusting phase term, the final sensing BF vector can be obtained as
\begin{align}
\w_{t,s}=\frac{g_c{g_s}^c}{|g_c{g_s}|}\bm{\beta}=\alpha\frac{g_c{g_s}^c}{|g_c{g_s}|}\A^\dag \D_v\p_{v,0} .
\end{align}
The underlying idea here is to make the two subbeams to add coherently at the desired direction (dominating communication direction). This approach is simple and demonstrates good performance in general. However, it is not optimized and may become less efficient particularly when communication and sensing subbeams locate closely in direction.

\subsubsection{Method 2: Joint Design }\label{sec-method2}
In this joint design method, we use a single desired multibeam waveform as the input to the ILS algorithm to get the combined BF vector directly. 
Referring to the ILS algorithm, one example of this method is described as follows.
\begin{enumerate}
\item[S1:] Obtain the BF vectors for communication and sensing subbeams separately and denote the magnitude vector of their radiation pattern as $\df_c$ and $\df_s$; 
\item[S2:] Let the magnitude of the desired BF waveform $\D_v$  be 
\begin{align}
\D_v=\diag(\max(\sqrt{\rho}\df_c,\sqrt{1-\rho}\df_s)).
\end{align}
That is, each diagonal element of $\D_v$ is the maximum of corresponding two elements in $\sqrt{\rho}\df_c$ and $\sqrt{1-\rho}\df_s$. Using the maximum instead of sum can better maintain the shape of the desired waveform and minimize the sidelobe.
\item[S3:] Use the ILS algorithm with the input $\D_v$ to get the BF vector $\w_t$.
\end{enumerate}

\subsubsection{Comparison of the Two Methods}

The two proposed methods have respective advantages and disadvantages, as can be observed from the discussions above and the simulation results to be presented in Section \ref{sec-simu}. The comparison is summarized below.

\begin{itemize}
\item Method 1, the separated design, can generally achieve higher BF gain for communication compared to Method 2. It is very flexible in adapting to varying requirements for power and pointing directions, as each BF vector can be updated individually before combination. The resulted variation in sidelobe is generally small and the BF gain at the desired directions can be generally maintained. When the communication  and sensing subbeams are very close in direction, however, they can form a beam with almost a single main beam. This may cause some wasted energy if there is no other significant multipath near the dominating multipath for communication;
\item Method 2, the joint design, provides excellent control for the shape of the BF waveform, but it generally achieves inferior BF gain as no phase alignment was able to be applied for the communication and sensing subbeams. It is also inflexible in BF updating, as individual subbeam cannot be changed in either power or pointing directions. Any change in power allocation or pointing direction requires the running of almost the complete process as discussed in Section \ref{sec-method2}. 
\end{itemize}

Based on the comparisons above, Method 1 provides better balance between system performance and complexity. We will mainly refer to Method 1 in the rest of this paper. Certainly combination methods optimized for some set receivers are yet to be developed.

\subsection{Updating the BF Vector $\w_t$}
Referring to Method 1, the major steps of generating and updating BF vectors can now be summarized as follows, in the case of without requiring to change the basic reference BF waveform.
\begin{enumerate}
\item[1.] Apply the phase shifting vector to shift the reference waveform to the desired pointing direction for both communication and sensing subbeams;
\item[2.] Apply Method 1 to combine the two subbeams to generate the multibeam.
\end{enumerate}

 \subsection{Receiver BF Design}\label{sec-rxbf} 
{ We briefly discuss how to design the receiver BF, corresponding to the transmitted multibeam signal here, for both active and passive sensing.}
\subsubsection{{Active Sensing in CTAS Stage} }
Considering that reflected multipath signals will generally come from the opposite of the transmit BF direction, the receiver scanning directions can be chosen accordingly. 

To exploit the transmit multibeam, we want to let the receiver BF to cover the directions of both the communication and scanning subbeams during each packet. This can be done either through using a receiver multibeam or a single beam over different time. In this paper, we consider the latter which is consistent with the sensing algorithm that will be presented in Section \ref{sec-sensing}. Since the direction of the communication subbeam is approximately unchanged during one complete cycle of scanning, we can distribute the receiver scanning directions surrounding the opposite direction of communication beam to the $N_t$ packets. In this case, one example of the receiver BF design is as follows: During each set of $N_r$ scanning directions in each packet, assign $N_r-1$ directions corresponding to the transmit sensing subbeam, and $1$ direction corresponding to the communication subbeam. Across all $N_t$ packets, there are a total of $N_t$ and $(N_r-1)N_t$ receiver BF scanning directions, corresponding to communication and sensing BF directions, respectively. These directions can be interlaced or overlapped, if desired. 

\subsubsection{{Passive Sensing in CRPS Stage}}
{For communication reception, the receiver BF needs to be fixed over one packet period, to make sure the channel is stable. Hence if passive sensing is implemented, the receiver can also use a multibeam, being fixed during one packet period, and corresponding to and varying with the transmit multibeam. Without using the direction-varying scanning beam per OFDM symbol as is used in active sensing, the angle resolution may be reduced. But the SNR can be improved by averaging the measurements over multiple OFDM symbols.}

{Note that for passive sensing, the time delay measurement has an unknown time offset term and hence the distance measurement is relative, as the transmitter and receiver are non-synchronized in timing clock. The directions and Doppler frequencies (or speeds) can be similarly measured to those in active sensing.}  

\subsection{{Comparison to Separated Communication and Sensing}}
{The advantages of using multibeam for joint communication and sensing over conventional separated communication and sensing such as time-division or frequency-channelization can be seen from the simple analysis below.}

{Refer to the time-division method where communication and sensing use separated timeslots and the ratio of the period between communication timeslot and the total timeslot is $a$. For the proposed multibeam scheme, we consider the case where the power distribution factor $\rho={a}$. For communications, we compare the Shannon capacities for the time-division and multibeam methods, which can be simplified as
\begin{align}
&C_{\text{MB}}=B\log2(1+a|h|^2 MP_t/\sigma_z^2),\ \text{for Multibeam},\notag\\
&C_{\text{TD}}=aB\log2(1+|h|^2  MP_t/\sigma_z^2),\ \text{for Time-division},
\end{align}
where $P_t$ is the total transmission power, $M$ is the number of antennas in the array, representing the maximal beamforming gain, $h$ is the propagation channel coefficient, $B$ is the bandwidth, and $\sigma_z^2$ is the noise variance. It is easy to verify that the ratio $C_{\text{MB}}/C_{\text{TD}}$ is always equal to or larger than $1$, with equality held when $a=1$. For example, for $a=1/2$, the ratio is $1.88$ and $1.94$ for $M|h|^2  P_t/\sigma_z^2$ being $5$ and $10$ dB, respectively. In the analysis above, we didn't even take into consideration the potential increasing of signal power via combining the signal from the sensing subbeam.}

{On the other hand, for sensing, using a simple example we can show that the multibeam method can perform at least as well as the time-division method. Athough the power of the scanning subbeam in the multibeam method is only $(1-a)$ of that in time-division, its scanning time is $1/(1-a)$ times longer. Therefore, in the multibeam method, just repeating the scanning at the same direction over the whole period and combining the received signals, we can already get the same received SNR for scanning with that in the time-division method. If we apply the mutual information analysis for sensing, we can also see that the mutual information for sensing in the multibeam method is always larger than that in the time-division method.}

\section{Sensing Algorithms in CTAS Stage} \label{sec-sensing}

The basic task in sensing is to determine distance, direction, and speed of environmental objects from received signals, and then extract information from these parameters. They can be obtained through both direct estimation of parameters $\tau_{\ell}$, $f_{D,\ell}$, $\theta_{t,\ell}$ and $\theta_{r,\ell}$, and data fusion processing over time, frequency and spatial domains. The main purpose of this section is to demonstrate that it is feasible to apply various algorithms, including low-complexity DFT (periodogram) method, and high resolution spectrum analysis and compressive sensing algorithms, complying with the proposed multibeam framework. {We will base our discussions on active sensing in the CTAS stage here. Most of the proposed algorithms can be extended to sensing in the CRPS stage straightforwardly.} 

For simplicity, we assume that the OFDM symbols used for sensing are indexed from $k=1$ and located in the same position in each packet. Let the packet period be $T_f$. 

Since the data symbols $\sfd$ is known to the receiver and $\w_t$ and $\w_r$ are fixed for at least one OFDM symbol, we can convert the received signal to the frequency domain and remove $\sfd$ via equalization. Ignore the variation of Doppler phase shift within one OFDM symbol and approximate it as a single value. For the $(k=(n_d-1)N_r+n_r)$-th OFDM symbol in the $n_t$-th packet we can get the frequency-domain channel estimate at the $n$-th subcarrier as 
\begin{align}
\hfi_{n,n_t,k}=&\sum_{\ell=1}^L \underbrace{(b_\ell(\w^T_t(n_t)\av(\theta_{t,\ell}))\w^T_r(n_r)\av(\theta_{r,\ell}))}_{g_{\ell}(n_t,n_r)} \cdot \notag\\
&\quad e^{-j2\pi n \tau_{\ell}f_0}e^{j2\pi f_{D,\ell}(kT_s+(n_t-1)T_f)}  + \tilde{z}_n/\sfi_n \notag\\
\label{eq-hfi}
\end{align}
where $\tilde{z}_n$ is the noise sample at the $n$-th subcarrier.

For fixed $\w_t$ and $\w_r$, $g_{\ell}(n_t,n_r)$ will be a function of $\ell$ and the signal in (\ref{eq-hfi}) presents a typical form of radar signal. {From (\ref{eq-hfi}), we can see that the maximum unambiguous resolution for delay $\tau_{\ell}$ and Doppler frequency $f_{D,\ell}$ are $\tau_\text{max}=1/f_0$ and $f_{D,\max}=1/T_s=f_0$. The maximum value for sensible range and moving speed can accordingly be determined. The granularity of the resolution (i.e., minimal resolvable values) for delay and Doppler mainly depends on the signal bandwidth and the measurement time period during which the channel parameters can be regarded as unchanged, respectively \cite{Sit11,Sturm11}. When high resolution algorithms such as those to be discussed in Section \ref{sec-highreso} are applied, the granularity of resolution also depends on other factors such as SNR and the number of observations.}

For the $n_t$-th packet, stack the signals $\hfi_{n,n_t,k}$ with the same $\w_r$ to a series of $N_r$ $N\times N_d$ matrices $\Yt(n_t,n_r)$, $n_r=1,\cdots,N_r$, where $(\Yt(n_t,n_r))_{n,n_d}=\hfi_{n,n_t,k}$. These elements are from all $N$ subcarriers over $N_d$ OFDM symbols, at an interval of $N_r$ symbols. In the noise-free case, we can get
\begin{align}
\Yt(n_t,n_r)=\C_{\tau}\D_g(n_t,n_r)\V_d,
\label{eq-yftr}
\end{align}
where 
\begin{align}
&(\C_{\tau})_{n,\ell}=e^{-j2\pi n\tau_\ell f_0};\notag\\
&\D_g(n_t,n_r)=\diag\{d_{\ell}(n_t,n_r)\}, \ \ell=1,\cdots, L,\notag\\
&\quad\quad d_{\ell}(n_t,n_r)\triangleq g_{\ell}(n_t,n_r)e^{j2\pi f_{D,\ell}(n_rTs+(n_t-1)T_f)};\notag\\ 
&(\V_d)_{\ell,n_d}=e^{j2\pi (n_d-1)N_rf_{D,\ell}T_s}.
\end{align}

\subsection{Low-resolution Low-complexity Approach: DFT Method}
We can apply simple spectrum techniques such as periodogram and 2D DFTs to each $\Yt(n_t,n_r)$ to get coarse estimates for $\tau_{\ell}$ and $f_{D,\ell}$ \cite{Braun14}. To generate a complete Delay-Doppler 2D-view, we can take the absolute values of the 2D-DFT outputs for all $(n_t,n_r)$, and then sum them up. 

We may want to average the measured signals to improve the SNRs or to synthesize over different scanning angles. However, such averaging operation needs to be cautious.

Note that adding together $\Yt(n_t,n_r)$ with different $n_r$s may not improve SNRs. Consider the case when $\w_r(n_t,n_r)$ generates a directional single beam pointing to the direction $\varphi_{n_t,n_r} $. We have  
\begin{align}
\w_r^T(n_t,n_r)\av(\theta_{r,\ell})=e^{j\pi (\sin(\theta_{r,\ell})-\sin(\varphi_{n_t,n_r}))(M-1)/2}\cdot\notag\\ \frac{\sin(\pi (\sin(\theta_{r,\ell})-\sin(\varphi_{n_t,n_r}))M/2)}{\sin(\pi (\sin(\theta_{r,\ell})-\sin(\varphi_{n_t,n_r}))/2)} .
\end{align}
Even when two pointing directions are within the range of the mainlobe of a single beam, the maximum phase difference can be close to $2\pi$.  This indicates that $d_{\ell}(n_t,n_r)$ could have quite different phase values for different $n_r$s.

However,  some columns in $\Yt(n_t,n_r)$ can be added together to improve SNRs. Recalling the example in Section \ref{sec-principle} where $f_{D,\ell}T_s\leq0.0072$, we can see that $N_rf_{D,\ell}T_s\ll 1$ for a typical $N_r=8$, but $N_rN_df_{D,\ell}T_s$ could be close to $1$. Hence such averaging is limited to approximately $4$ consecutive columns in $\Yt(n_t,n_r)$s. 

One major problem with this method is its limited resolution capability, particularly in resolving the Doppler frequencies. The Doppler frequencies are typically quite small due to limited moving speed. Resolution can be improved by using larger $N_rN_d$ values, which requires longer packets.

\subsection{High-resolution Approach: 1D Compressive Sensing}\label{sec-highreso}
Resolution for $\tau_{\ell}$ and $\f_{D,\ell}$ may be improved by using traditional super-resolution spectrum analysis techniques such as 2D-ESPRIT \cite{Sahnoun17} and 2-D Matrix Pencil \cite{yhua92, Chen07}, based on the signal model in (\ref{eq-yftr}). However, these techniques require the number of measurements to be larger than the number of multipath signals in each dimension. For the parameter $N_d$, it is hard to be satisfied. It is also hard for these methods to combine the measurements from multiple angles to improve the estimates for delay and Doppler frequency. 

More recently, compressive sensing (CS) techniques have been widely applied in radar imaging \cite{Herman09, Hadi2015}. The four parameters, AoDs, AoAs, delay and Doppler frequency in (\ref{eq-hfi}) can be estimated either individually or jointly by forming from 1D to 4D CS models \cite{Hadi2015}. {Generally, higher-dimension models can lead to better performance, but they will also require reconstruction algorithms with much higher complexity. On the other hand, current CS techniques mostly work on on-grid (or quantized) signal models. There exist some techniques dealing with off-grid models, such as the perturbation approach \cite{Fannjiang13} and atomic norms \cite{Tang13}, but they are currently limited to low-dimension ($\leq 2$) problems, and also have respective constrains on the parameter estimation range and the minimum separation of the parameter values. CS techniques for high-dimension off-grid models are still immature.} In our framework, quantization errors in the on-grid model can be quite large for Doppler frequency, AoA and AoD due to the limited number of measurements in these domains. This makes current high-dimension CS solution ineffective here. 

In this paper, we propose a novel estimation algorithm based on 1D multiple measurement vector (MMV) CS techniques. {For clarity and also considering the maturity of technology, we will base our algorithm on the on-grid model, which will be shown to work well even for continuous parameters in the simulation results in Section \ref{sec-simu}. Existing off-grid-model based techniques such as \cite{Fannjiang13, Tang13}  can also be applied when being extended to the MMV setup.} We use MMV-CS to estimate delays, and through the estimated delays to estimate the rest parameters. The quantization error for delay is small enough for achieving satisfactory estimation performance, because both the number of measurements $N$ (number of subcarriers) and  the bandwidth are sufficiently large. Using this algorithm we can also combine all measurements across the whole $N_t$ packets, if desired, to get the delay estimates with significantly improved SNR. This CS based method also provides great flexibility in the receiver scanning BF design. For example, it is possible to use multibeam receiver BF, which is to be investigated in the future. 

This algorithm essentially exploits the fine accuracy of the amplitude estimates in the MMV model. Other parameters such as Doppler frequency can then be estimated from the amplitudes. This will separate multipaths to multiple delay bins, which will generally make the number of multipath in each delay bin to be smaller than $N_d$ so that 1-D spectrum analysis techniques can be applied. Conversely, the amplitude estimation in conventional spectrum analysis techniques are less accurate, and hence cannot be used for estimating Doppler frequencies directly. 

\subsubsection{Estimation of Delays $\tau_{\ell}$}
In order to estimate $\tau_{\ell}$, we can treat the product of $g_{\ell}(n_t,n_r)$ and the Doppler phase term in (\ref{eq-hfi}) as varying amplitude over different measurements for the delays, and then formulate it as an MMV CS problem.  

Modifying from (\ref{eq-yftr}) and ignoring the quantization error, we can get a delay-quantized on-grid model for the measurements from $N_d$ OFDM symbols as 
\begin{align}
{\Yt(n_t,n_r)}=\Cq_{\tau}\underbrace{\Pc(n_t,n_r)\D_g(n_t,n_r)\V_d}_{\triangleq\U(n_t,n_r)},
\label{eq-ytqt}
\end{align}
where $\Cq_{\tau}$ is a $N\times L_p$ dictionary matrix with elements
\begin{align}
(\Cq_{\tau})_{n, q}=e^{-j2\pi n q/L_p},
\end{align}
corresponding to the quantized delay $qN/(BL_p) $, and $\Pc(n_t,n_r)$ is an $L_p\times L$ matrix with elements either zeros or ones, associating the multipath delays with the elements in $\D_g(n_t,n_r)\V_d$. 

Note that we can use an overcomplete dictionary $\Cq_{\tau}$ to reduce quantization error. In this case, we get an interpolated DFT matrix with dimension $N\times L_p$. We found that $L_p=2N$ is a good choice that balances the quantization error and the correlation between different columns of the dictionary matrix. 

We consider three cases according to the number of multipaths at each delay at each receiver beamforming scanning direction: (1) No multipath has delay value $qN/(BL_p) $, (2) A single multipath has delay $qN/(BL_p) $, and (3) Multiple multipaths have the same delay $qN/(BL_p) $ but different Doppler frequencies. For the three cases, the  $q$-th row 
of $\Pc(n_t,n_r)$ has all zeros, a single 1 or multiple 1's, respectively. The total number of 1's in $\Pc(n_t,n_r)$ equals to $L$. Therefore the $\{q,n_d\}$-th element of $\U(n_t,n_r)$ is given by
\begin{align}
&u_{q,n_d}(n_t,n_r)\notag\\ 
=&\left\{
\begin{array}{ll}
0, & \text{Case 1;} \\ 
d_\ell(n_t,n_r) e^{j2\pi (n_d-1)N_rf_{D,\ell}T_s}, & \text{Case 2;}\\
\sum_{\ell\in\textit{ S}}d_\ell(n_t,n_r)e^{j2\pi (n_d-1)N_rf_{D,\ell}T_s}, & \text{Case 3.}
\end{array} 
\right. 
\label{eq-uqnd}
\end{align}
where $\textit{S}$ is a set with more than one multipath indexes.

When organizing measurements into an MMV mathematical model, we want to make sure they have the common support (delay values), otherwise no improvement on estimation performance can be achieved. Equation (\ref{eq-ytqt}) forms a basic MMV signal model where $N_d$ measurements are used. For receiver sensing beams with close scanning directions or even with overlapped beams, they may see multipaths from a single large object or a cluster of objects with the same quantized delay. Hence it will be beneficial to stack them into one MMV model. We consider two combinations here: (1) combine the $N_r-1$ receive BF results corresponding to the transmit sensing subbeam direction in each packet, and (2) combine each receive BF result corresponding to the transmit communication subbeam direction in a packet across $N_t$ packets. We represent the extended MMV signal model as
\begin{align}
\Rt(n_t)&=[\Yt(n_t,1),\ldots,\Yt(n_t,N_r-1)]\notag\\
&=\Cq_{\tau}\underbrace{[\U(n_t,1),\ldots,\U(n_t,N_r-1)]}_{\U}
\label{eq-case1}
\end{align}
for Combination 1, and 
\begin{align}
\Rt(n_t)&=[\Yt(1,N_r),\ldots,\Yt(N_t,N_r)]\notag\\
&=\Cq_{\tau}\underbrace{[\U(1,N_r),\ldots,\U(N_t,N_r)]}_{\U}
\label{eq-case2}
\end{align}
for Combination 2.  

For each $\Rt(n_t)$, we can then apply some MMV CS algorithm such as MMV Bayesian Compressive Sensing (BCS) \cite{Wipf07, zhilin11} or MMV OMP \cite{Ding12} to get estimates for $\U$. Once $\U$ is known, the quantized delay and Doppler frequencies can then be estimated from the non-zero rows in $\U$, referring back to individual composing matrix in $\U$. 

Using Combination 1 in (\ref{eq-case1}) as an example, we now study how to extract estimates for delays, and their associated Doppler frequencies and AoAs.  

Referring to (\ref{eq-uqnd}), the cross-correlation between two neighbouring elements in the $q$-th row of $\U(n_t,n_r)$ can be computed as
\begin{align}
&\lambda_{q,n_d}(n_t,n_r)=u^*_{q,n_d}(n_t,n_r)u_{q,n_d+1}(n_t,n_r)\notag\\
=&\left\{
\begin{array}{ll}
0, & \text{for Case 1;} \\ 
|g_\ell(n_t,n_r)|^2 e^{j2\pi N_rf_{D,\ell}T_s}, & \text{for Case 2};\\
\sum_{\ell_1\in\textit{ S}}\sum_{\ell_2\in\textit{ S}}d_{\ell_1}(n_t,n_r)d_{\ell_2}(n_t,n_r)\cdot & \\
\qquad e^{j2\pi N_r(n_df_{D,\ell_1}-(n_d-1)f_{D,\ell_2})T_s}, & \text{for Case 3.}
\end{array}\right.
\label{eq-lambda}
\end{align}

Let the mean and variance of $\{\lambda_{q,n_d}(n_t,n_r)\}, n_d=1,\ldots, N_d-1$ be 
\begin{align}
\bar{\lambda}_q(n_t,n_r)&=\frac{1}{N_d-1}\sum_{n_d=1}^{N_d-1}\lambda_{q,n_d}(n_t,n_r),\notag\\
\sigma_{q,n_d}(n_t,n_r)&=\frac{1}{N_d-1}\sum_{n_d=1}^{N_d-1}(\lambda_{q,n_d}(n_t,n_r)-\bar{\lambda}_q(n_t,n_r))^2.\notag
\end{align} 

Based on (\ref{eq-lambda}), we can now decide whether there are multipath signals at delay $qN/(BL_p) $ by comparing $\bar{\lambda}_q(n_t,n_r)$ with a \emph{delay-dependent threshold} $\eta_q$: If $\bar{\lambda}_q(n_t,n_r)\geq \eta_q$, there is at least one multipath; otherwise, no multipath. The threshold $\eta_q$ can be set according to the anticipated power of the received signal for the delay $qN/(BL_p) $. 

\subsubsection{Estimation of Doppler Frequency}
Without differentiating between Cases 2 and 3, we can apply conventional spectrum analysis techniques such as 1D ESPRIT to find the estimates or Doppler frequencies from each $u_{q,n_d}(n_t,n_r)$, referring back to (\ref{eq-uqnd}), The number of resolvable multipaths will be equal to or smaller than $N_d/2$. 

We can use another method with much lower complexity to find Doppler frequencies if there is only one multipath in each scanning direction and each delay bin (i.e., Case 2). This could be a common case when the obstacle can be treated as a point source. To filter out possibly a limited number of Case 3, we need a mechanism to differentiate between Cases 2 and 3. We tried the method that compares $\sigma_{q,n_d}(n_t,n_r)$ with a second threshold, based on the fact that $\sigma_{q,n_d}(n_t,n_r)=0$ in the noiseless case in Case 2. However it seems not working reliably and hence it remains as an open problem yet to be solved. 

In Case 2, the Doppler frequency can be directly estimated from the angle of $\lambda_{q,n_d}(n_t,n_r)$
\begin{align}
\widehat{f_{D,\ell}}=\angle(\lambda_{q,n_d}(n_t,n_r))/(2\pi NrT_s).
\end{align}
Since $2\pi Nrf_{D,\ell}T_s\ll 1$, there could be only one possible value for $\f_{D,\ell}$.

We note that for Combination 2, similar process applies, but it could be better, from the combination, to exclude $\U(n,N_r)$s obtained when communication and sensing subbeams are close in directions, as $d_\ell(n_t,N_r)$ in these cases can be quite different to the others due to the significant contribution from the sensing subbeam.

\section{Simulation Results}\label{sec-simu}

We present simulation results to validate the proposed framework, considering a system with major communication parameters detailed in Section \ref{sec-principle}. Each packet is assumed to have 60 OFDM symbols. Node A has two 16-element ULAs, and the interval between antenna elements is half wavelength. Node A moves at speed 20 m/s, and obstacles (objects in surrounding environment) have random speed between -40 and 40 m/s. 

The communication direction, where Node B locates, is assumed to be at 0 degree. Assume there is a dominating LOS multipath between Nodes A and B. The mean power ratio between this LOS and the rest multipath signals is 10dB. All the communication multipaths are uniformly distributed within a direction range of 33 degrees. 

\subsection{Multibeam Generation}

Jointly considering the number of OFDM symbols in one packet and the size of the array, we set $N_t=8, N_r=5$, $N_d=12$, where $N_d$ is set a large value to achieve better estimation for Doppler frequencies. {In this case, from the other system parameters as provided in Section \ref{sec-principle}, we can see that a complete sensing cycle lasts about $0.768ms$, which is also the communication transmission period. Transmission within this cycle does not have to be continuous, that is, the $N_t$ packets does not have to be transmitted continuously at a time. The transmitter can also work in the receiving mode for communication. These parameters can be readily adapted to the actual  TDD timeslot allocation in the communication protocol.} 
	
Based on these parameters, we want to generate transmit BF waveforms that cover a scanning range from -60 to 60 degrees, overlapped at approximately 3dB beamwidth. The equivalent scanning directions are hence spaced at 0.2, leading to the desired actual pointing directions of the 8 scanning subbeams to approximately $-54.3, -37.8, -24.4, -12.3, 10.8, 22.8, 35.9$ and $51.9$ degrees. {The size of the matrix $\A$ is $M\times K$, with $M=16$ and $K$ selected as $160$ to illustrate smooth curves for BF waveforms. In real applications, a much smaller $K$ such as $K=5M$ can be applied.} Note the non-uniform actual scanning directions is purely resulted from the requirement of applying the simple displaced BF waveform generation method.

\begin{figure}[t]
\centering
\includegraphics[width=0.9\columnwidth]{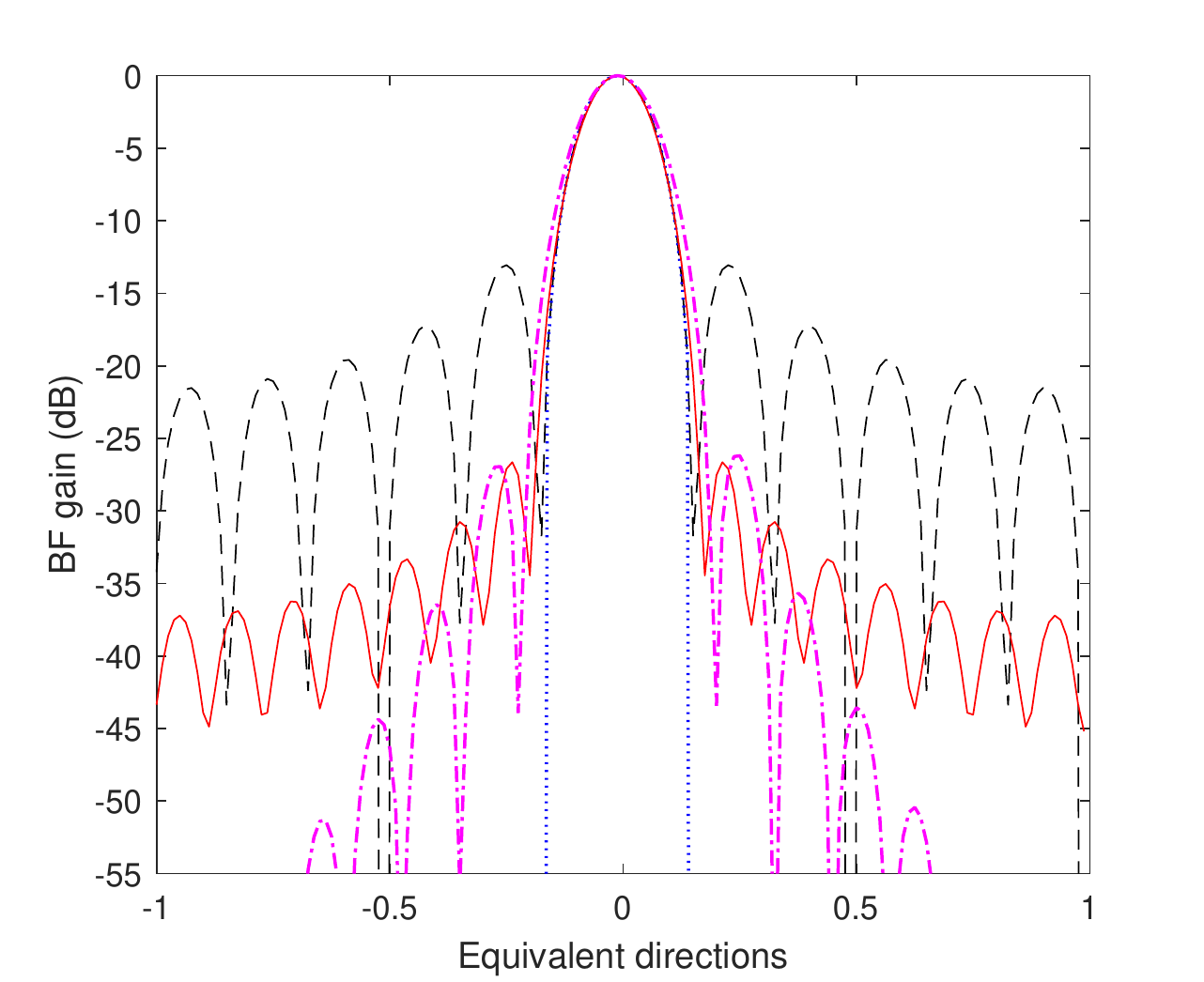}
\caption{Reference sensing subbeam, generated based on the radiation pattern of a 12-element ULA. Black-dashed curves. Please refer to the third paragraph in Section VI.A for detailed descriptions for different curves. }
\label{fig-refbeam}
\end{figure}

In Fig. \ref{fig-refbeam}, we show the basic reference BF waveform for the sensing subbeam. The black dashed curve is the radiation pattern of a conventional 12-element ULA. The blue dotted curve, which takes the mainlobe of the dashed curve and sets the rest to zeros, is the desired magnitude input to the ILS algorithm. {The red solid curve represents the output from the ILS algorithm with $\D=\I$ and is the reference BF waveform used for generating the scanning BF waveforms. The pink dashed-dotted curve represents the output waveform from the ILS algorithm with $\D$ being a special diagonal matrix. Its diagonal elements correspond to the normalized version of an exponential function $\exp([80, 79, \cdots,0, 1,\cdots, 79]/15)$. Note that the reference waveform has already achieved much lower sidelobes than the initial radiation pattern (black dashed curve), which is a desired property. The pink-dotted curve shows even lower sidelobes, which is achieved by applying this special weighting matrix $\D$. This clearly demonstrates an important usage for  $\D$: obtaining BF vectors that can generate lower sidelobes, than those achievable by only setting the sidelobe of the desired magnitude response to zero. Through $\D$, we apply different weights to the square errors as shown in (\ref {eq-ls}), and can obtain different similarities for different segments of the waveform between the generated BF waveform and the desired one.} The communication subbeam is similarly generated, but uses the radiation pattern from a 16-element ULA to obtain higher gain. 

\begin{figure}[t]
\centering
\includegraphics[width=0.9\columnwidth]{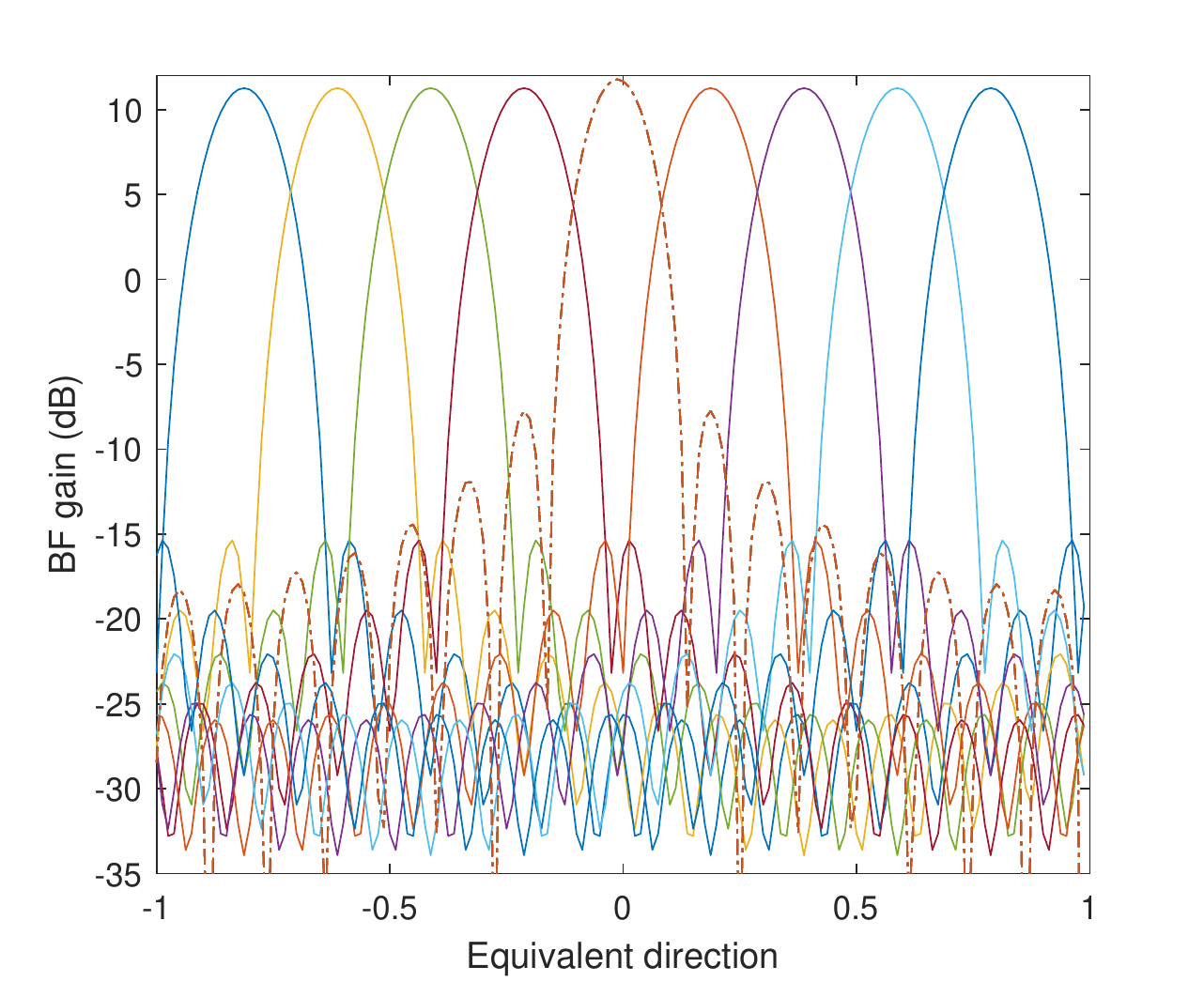}
\caption{Communication and displaced sensing subbeams plotted over equivalent directions. All the sensing subbeams are generated using the method described in Section \ref{sec-displace}, from the reference sensing subbeam at direction 0, shown in Fig. \ref{fig-refbeam}.}
\label{fig-indi}
\end{figure}

Fig. \ref{fig-indi} plots the 8 individual sensing subbeams together with the communication subbeam before combination, generated via the displaced BF waveform method. The communication subbeam has a narrower mainrobe and higher BF gain as it is generated from the radiation pattern of a 16-element ULA. Note that these sensing subbeams have the same main beamwidth and are spaced equally over the equivalent directions.

\begin{figure}[t]
\centering
\includegraphics[width=1\columnwidth]{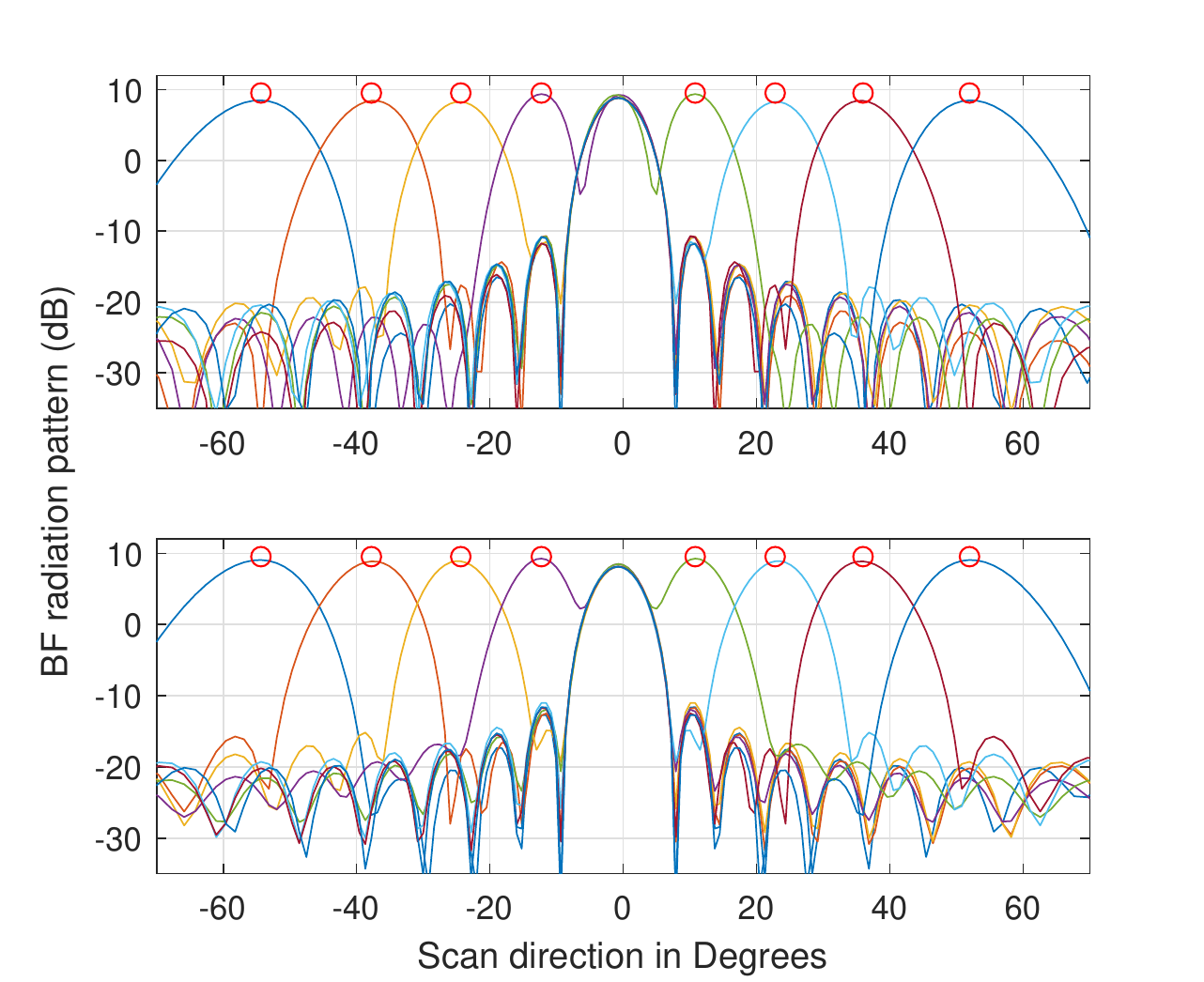}
\caption{Eight combined BF waveforms plotted together, over actual directions in degrees, for Method 1 (top subfigure) and 2 (bottom subfigure).}
\label{fig-multibeam}
\end{figure}

\begin{figure}[t]
\centering
\includegraphics[width=1\columnwidth]{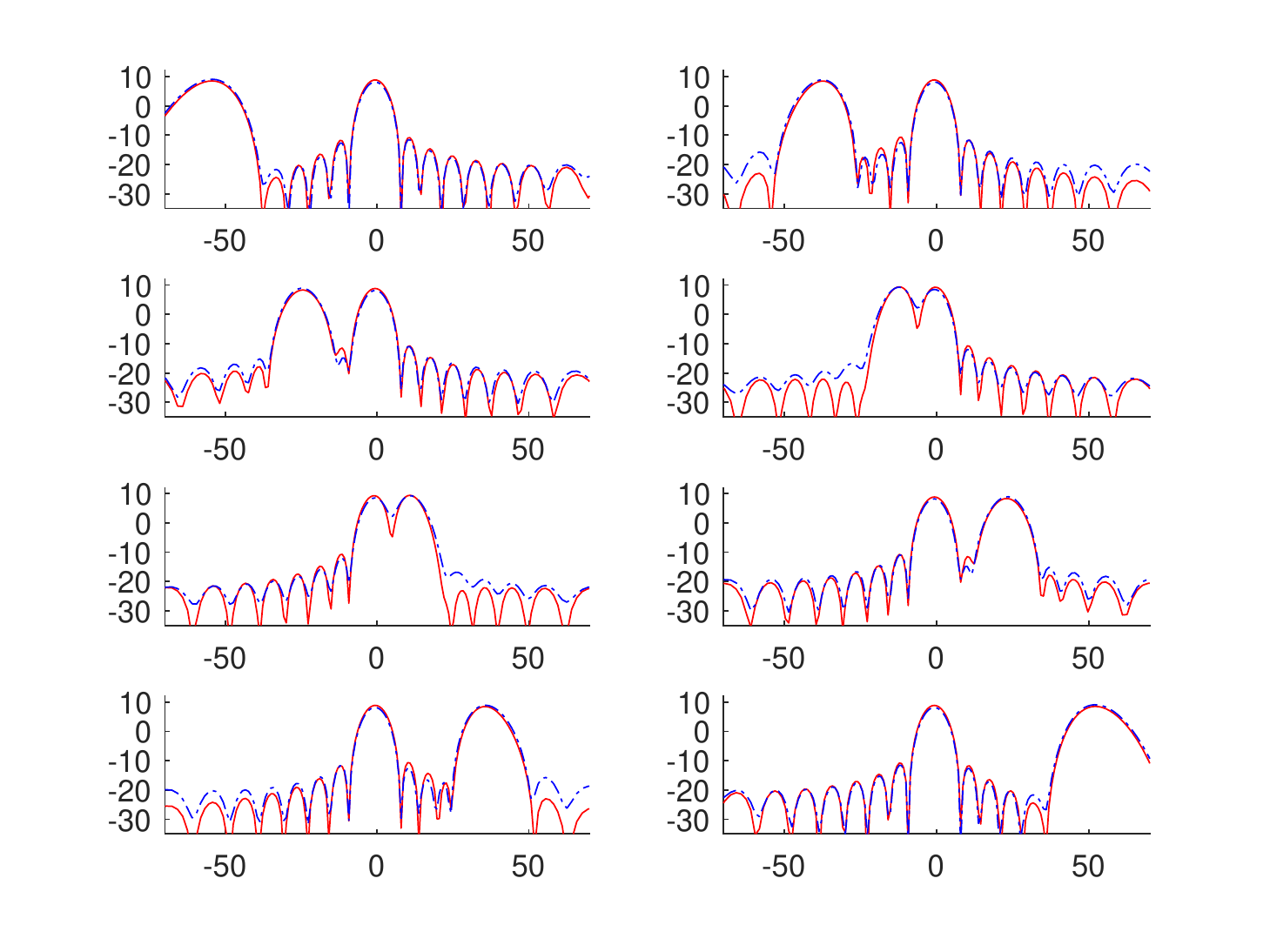}
\caption{Individual BF waveform used over each of $N_t$ packets for Method 1 (in red solid curve) and 2 (in blue dashed curve).}
\label{fig-indi2}
\end{figure}

In Fig. \ref{fig-multibeam}, we present the 8 BF waveforms after combining communication and sensing subbeams using Methods 1 and 2. Fig. \ref{fig-indi2} plots individual BF waveform for the two methods. We can see that Method 2 has better control on the overall BF waveform shape and gains. Here the x-axis is the actual directions. The figure shows that the beamwidth increases with increased distance to direction at 0 degree (that the reference BF waveform points to). Narrower beamwidth can generally achieve better resolution in direction. Hence this result matches well with the scanning requirements in practical applications, i.e., regions closer to Node A in direction is more of interest.

\begin{figure}[t]
\centering
\includegraphics[width=1\columnwidth]{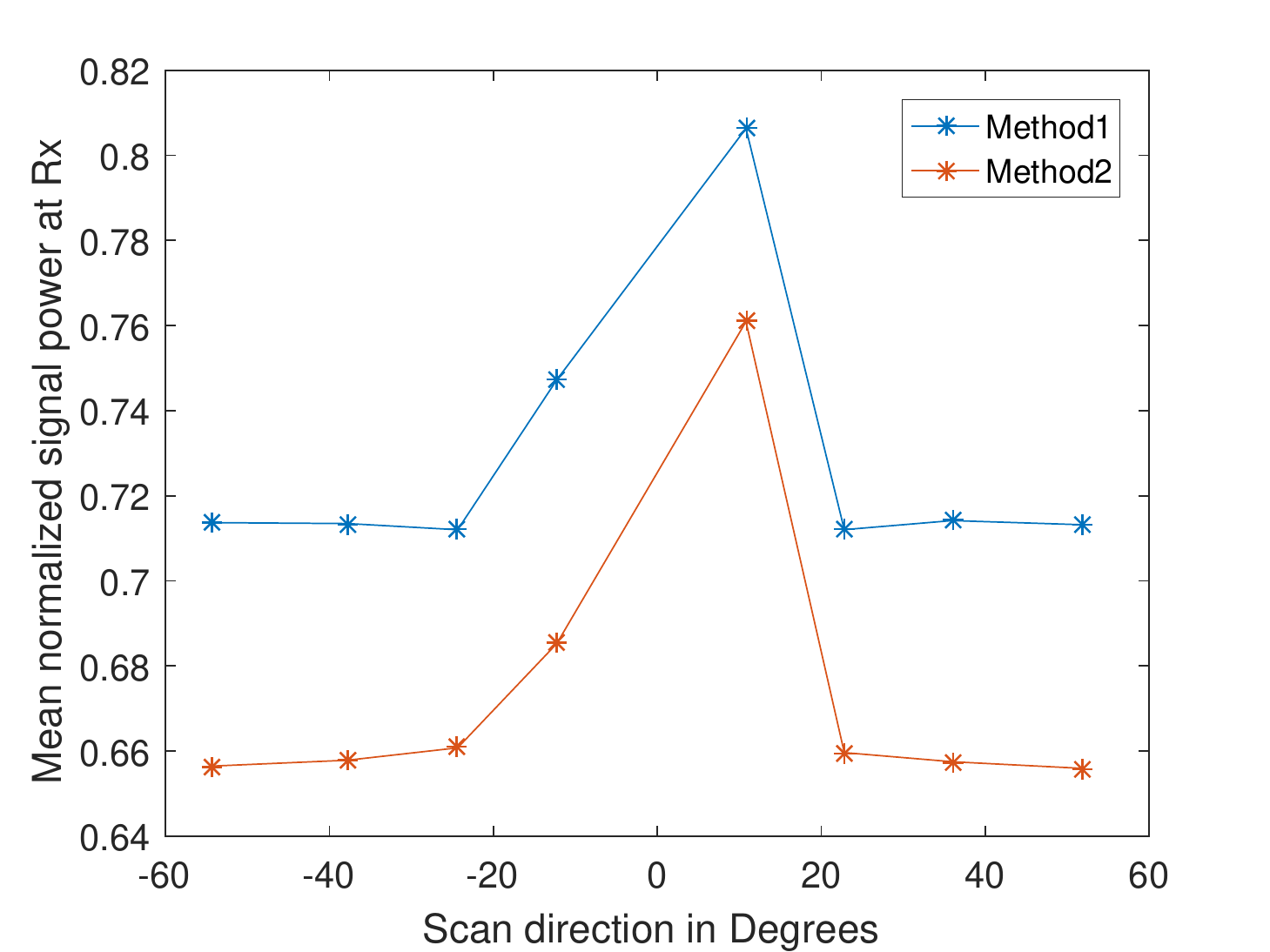}
\caption{Averaged normalized signal power at the receiver for Method 1 and 2, normalized to the received power when a single beam is used for communication only.}
\label{fig-gain}
\end{figure}

Fig. \ref{fig-gain} compares the averaged received signal power for the two subbeam combining methods, where the transmitting communication subbeam points to the dominating AoD and a maximal ratio combining receiver is used. The power is normalized to the value obtained when a single beam pointing to the dominating AoD is used for communication only. We can see that compared with Method 2, Method 1 achieves approximately $6\%$ increase at various scanning directions thanks to the use of ``coherent'' combining phase.

\subsection{Sensing Results}
To test the sensing performance in the proposed multibeam framework, we use the following additional system configurations. We generate a scattering environment with 12 scatters, and each scatter is assumed to be a point source in terms channel propagation. No radar cross-section information is assumed and this example is purely based on free-space pathloss.  We assume a pathloss model with path-loss factor $4$ for sensing. The transmission power is $25$ dBm, and the total thermal noise in the receiver is $-94$ dBm. In all the figures here, red circles are for values of the actual obstacles.

The obstacles/scatters are uniformly distributed over a distance up to 30 m, and over a range of AoAs from -60 to 60 degrees. All values of distance, moving speed and AoA are off-grid and continuous. 

For CS, the interpolated DFT matrix is used as the dictionary with an interpolation factor of 2. This enables the application of fast Fourier transform in the CS algorithm. We use BCS algorithm to solve the MMV problem.

\begin{figure}[t]
\centering
\includegraphics[width=1.1\columnwidth]{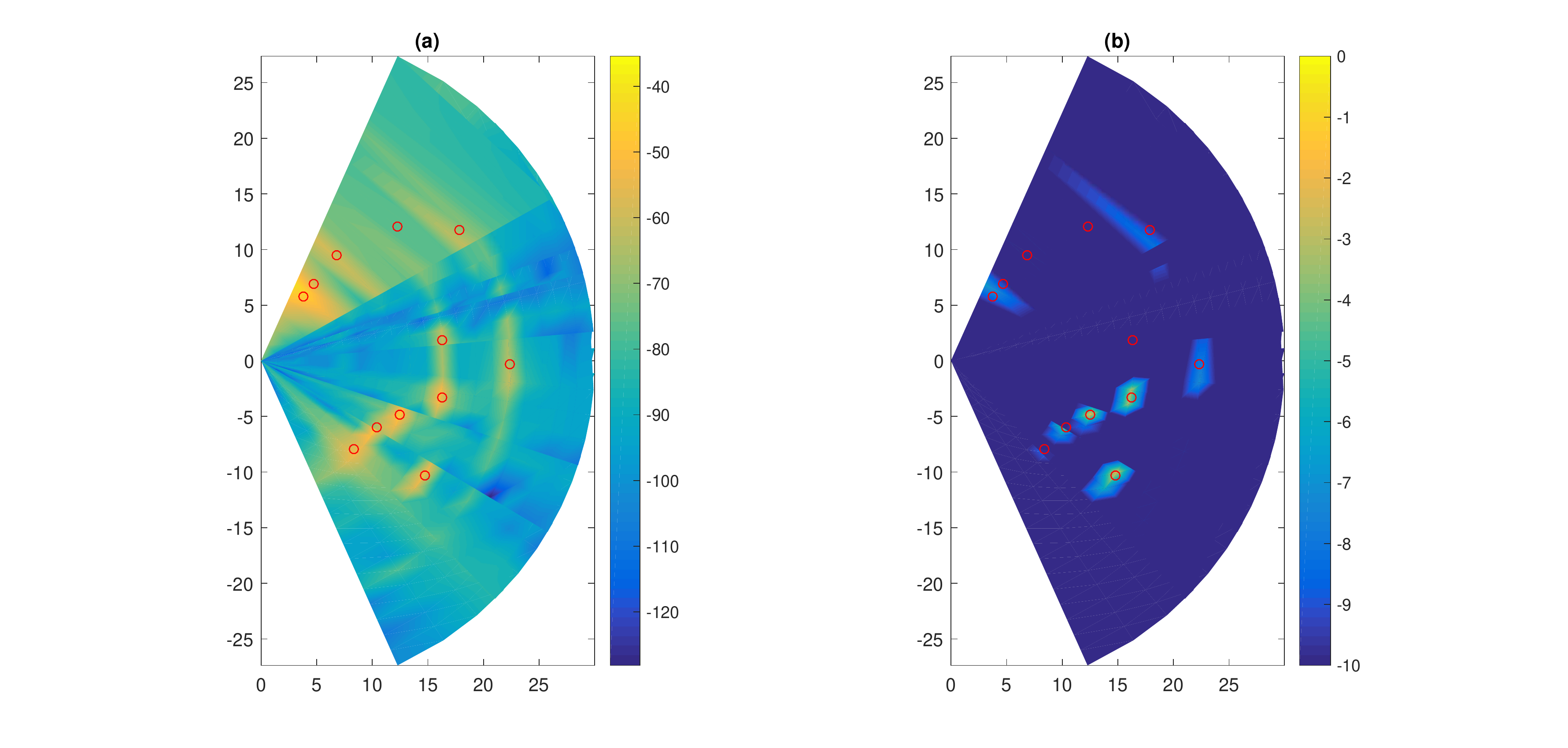}
\caption{Plots of distance (in meter) and AoA (in degrees) estimation results in polar coordinates using 1D IDFT: (a) Direct plot of power in dB; and (b) post-processed. }
\label{fig-idft}
\end{figure}

In the 2D DFT estimation, Doppler frequencies are mixed in the first bin and cannot be resolved. Hence in Fig. \ref{fig-idft}, we only present distance-AoA estimation results with the application of 1D IDFT. Fig. \ref{fig-idft}(a) shows the directly obtained results, while Fig. \ref{fig-idft}(b) shows the results after post-processing: Apply a pathloss model to compensate the power for IDFT outputs at different delays (corresponding to different distances), normalize the compensated estimates to their maximum, and set those estimate smaller than -10 dB to -10 dB (or much lower for higher sharpness in the plot). After the post-processing, locations of the obstacles become much clearer than those in Fig. \ref{fig-idft}(a), but are still not very accurate due to the low-resolution nature of the DFT algorithm. 

\begin{figure}[t]
\centering
\includegraphics[width=1.2\columnwidth]{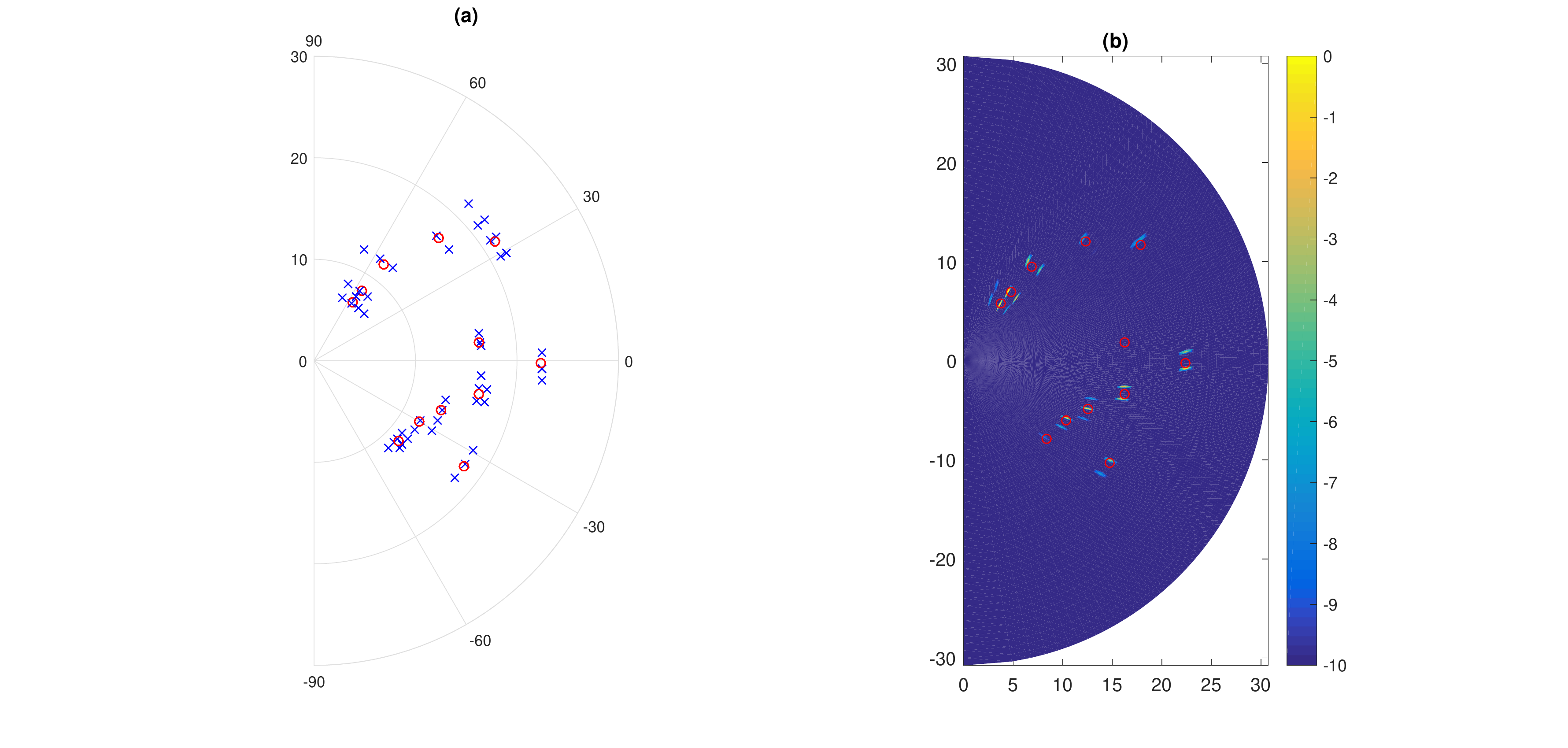}
\caption{Distance (meter) and AoA  (degrees) estimation results using the proposed CS algorithm: (a) Direct plot of the location estimates; and (b) post-processed results including both location and normalized power of the estimates in dB.}
\label{fig-cs}
\end{figure}

In Fig. \ref{fig-cs}, we show the distance-AoA estimation performance for the proposed CS algorithm: the direct location estimates in (a), and the location and power of the estimates after a similar post-processing with that in Fig. \ref{fig-idft}(b). Despite of some scattered estimates around the true locations, the figure demonstrates significantly improved resolution and accuracy in the distance-AoA estimation, compared to the DFT algorithm.
 
\begin{figure}[t]
\centering
\includegraphics[width=1.1\columnwidth]{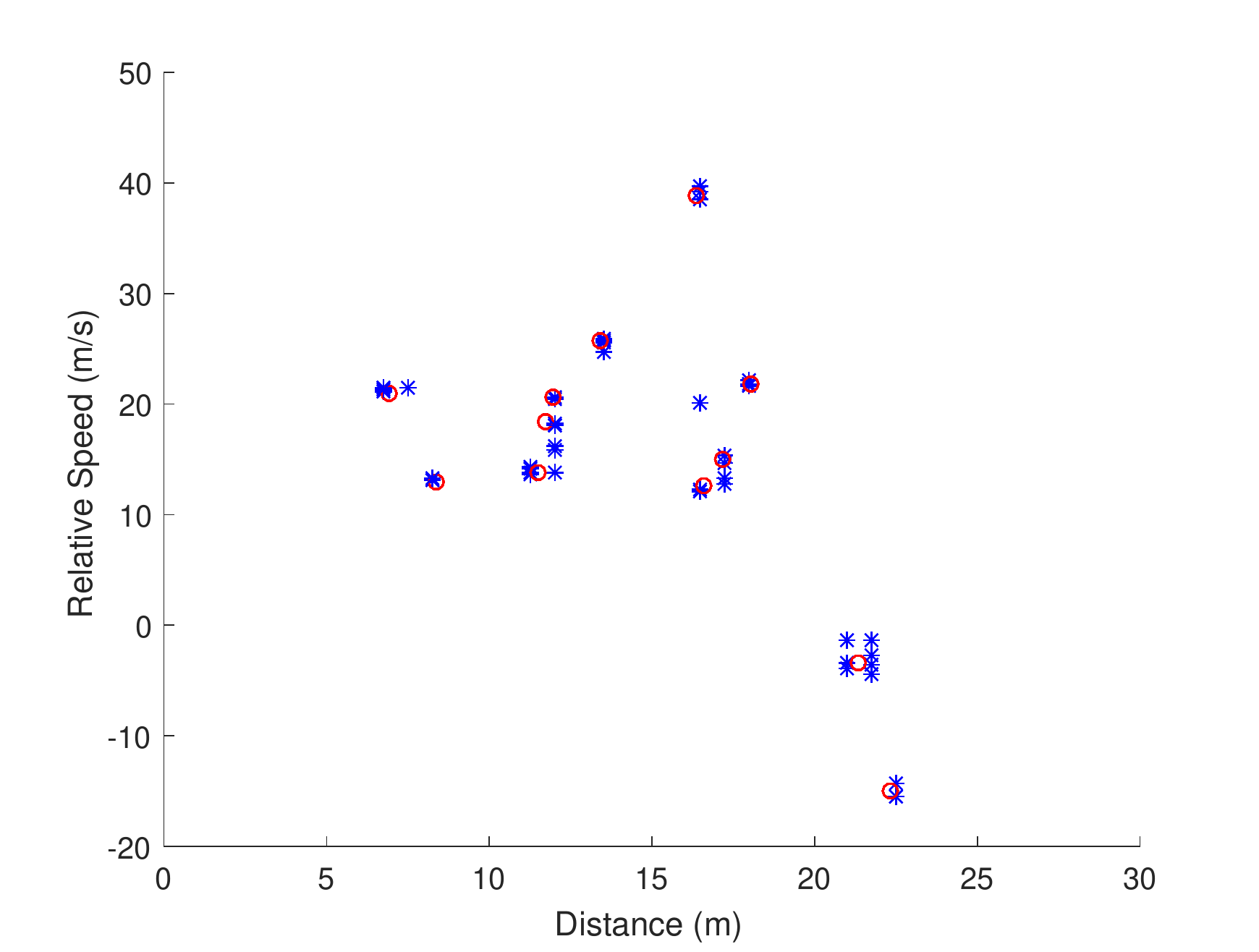}
\caption{Estimation of the relative speeds (m/s) versus estimated distance. }
\label{fig-speed}
\end{figure}

Fig. \ref{fig-speed} presents the associated estimates for the relative speed, aligned to the distance estimates. Estimates further away from the true values are still visible, but most estimates achieve good accuracy.

\section{Conclusions}

We have presented a novel multibeam framework with steerable analog antenna arrays for joint communication and sensing, which is very promising for platforms such as smart cars and UAVs. We proposed system architecture, protocols, BF design, multibeam generation and updating, and sensing parameter estimation algorithms for this framework. We demonstrated that using the proposed framework, it is feasible to seamlessly integrate sensing into standard TDD packet communication systems with OFDM modulation. Simulation results validate the effectiveness of the proposed system, multibeam generation and sensing algorithms. 

This is only the first step in exploring the strong potentials of the multibeam technology in JCAS. There are many challenging problems and possible improvements yet to be done, for example, multibeam BF vector generation with quantized magnitude and phase values, communication and sensing subbeam combination methods optimized with respect to certain criterion, and sensing algorithms that {work for high-dimension and off-grid models} and that can resolve AoAs and AoDs beyond the conventional concept of scanning.

\bibliographystyle{IEEEtran}
%

\end{document}